\documentclass{aa}
\usepackage{amsmath}
\usepackage{graphicx}
\usepackage{txfonts}
\usepackage{natbib}
\bibpunct{(}{)}{;}{a}{}{,} % to follow the A&A style
\newcommand{\teff}{\mbox{${T}_{\rm eff}$}}

\newcommand{\rsol}{\mbox{${\rm R}_{\odot}$}}
\newcommand{\msol}{\mbox{${\rm M}_{\odot}$}}
\newcommand{\msun}{\mbox{${\rm M}_{\odot}$}}
\newcommand{\lsol}{\mbox{${\rm L}_{\odot}$}}

\newcommand{\simgt}{\lower.5ex\hbox{$\; \buildrel > \over \sim \;$}}
\newcommand{\simlt}{\lower.5ex\hbox{$\; \buildrel < \over \sim \;$}}

\begin{document}
\title{Constraining fundamental stellar parameters using seismology}
\subtitle{Application to $\alpha$ Centauri AB}

\titlerunning{Constraining fundamental stellar parameters using seismology}

\author{A. Miglio \and J. Montalb\'an}
\institute{Institut d'Astrophysique et de G\'eophysique de l'Universit\'e de Li\`ege,
All\'ee du 6 Ao\^ut, 17 B-4000 Li\`ege, Belgium}

\offprints{A. Miglio,\\ \email{miglio@astro.ulg.ac.be}}

\abstract{We apply the Levenberg-Marquardt minimization algorithm to seismic and classical observables of the 
$\alpha$Cen binary system in order to derive the fundamental parameters of $\alpha$CenA+B, and to analyze the
dependence of these parameters on the chosen observables,
on  their uncertainty, and on the physics used in stellar modelling. The seismological data are those by 
\cite{Bouchy02} for
$\alpha$Cen~A, and those by \cite{Carrier03} for $\alpha$Cen~B.
We show that while the fundamental stellar parameters do not depend on the treatment of convection adopted 
(Mixing Length Theory -- MLT -- or ``Full Spectrum of Turbulence'' -- FST), the age of the system depends 
on the inclusion of gravitational settling, and is
 deeply biased by the small frequency separation  of component B.  
 We try to answer  the question of the universality of the mixing length parameter, and we find a
statistically reliable  dependence of the $\alpha$--parameter on the HR diagram location 
 \citep[with a  trend  similar to the one predicted by ][]{Ludwig99}.
  We propose the frequency separation ratios by \citet{Roxburgh03}
as  better observables to determine the fundamental  stellar parameters, and to use the large frequency 
separation and 
frequencies to extract information about the stellar structure. The effects of diffusion, and equation of state 
on the oscillation frequencies  are also studied, but present seismic data do not allow their detection.
\keywords{stars: oscillations -- stars: interiors -- stars: fundamental parameters -- stars: individual: 
$\alpha$ Cen}}
\maketitle
%________________________________________________________________

\section{Introduction}
$\alpha$~Cen~AB  is the binary system closest to the Earth
(d=1.34pc). It shows an eccentric orbit (e=0.519) with a period of
almost 80 years \citep{Pourbaix02}.
% Actually $\alpha$~Cen is a triple star system.
%The third member of the
%system, $\alpha$~Cen~C or Proxima $\alpha$~Cen, is a M5.5 Ve flare star (V=11.05) about 12000 AU distant from
%$\alpha$~Cen and only d=1.29 pc from the Sun (Perryman et al., 1997).
$\alpha$Cen~A is a G2V star and $\alpha$Cen~B
 a K1V one, sligthly hotter
and cooler respectively than the Sun.
 Thanks to the high  apparent
brightness (V$_{\rm A}=-0.01$ and V$_{\rm B}=1.33$) and to the large parallax of the components, their
stellar  parameters are among the best known of any star except the Sun.
The binarity, their well determined characteristics and the solar-like oscillations  detected in
both stars, provide a unique opportunity to test our knowledge on stellar evolution in conditions
slightly  different from the solar one.

As a consequence, a great number of theoretical studies dealing with $\alpha$Cen has been published since
the one by \citet{Flannery78} (see \citet{Eggenberger04} for a comprehensive review).
Before the definitive identification of p-mode frequencies in the $\alpha$Cen~A power spectrum by
 \citet{Bouchy02}, the uncertainty in the parallax (and therefore in the masses) and in the chemical
composition did not allow an unambiguous determination of stellar parameters. Some controversial results
came up concerning, for instance, the universality  or not of the mixing-length parameter ($\alpha$)
describing the stellar convection \citep{Noels91,Edmonds92,Lydon93,Neuforge93,Morel00,Fernandes95,Guenther00}; 
the role of  the chemical composition on discriminating between
 these two possibilities \citep{Fernandes95}, and on the presence or not of a convective core,
 and its effect on the age of the system \citep{Guenther00}. Some efforts \citep{Brown94,Guenther00,Morel00} 
 were also devoted to study the capability of solar-like
oscillations expected in $\alpha$Cen  \citep{Kjeldsen95} to constrain the fundamental stellar
parameters and the physics included in stellar models.

In addition to the p-mode identification by \citet{Bouchy02},   \citet{Pourbaix02}
improved the precision of the orbital parameters and, adopting the parallax derived by
 \citet{Soderhjelm99}, provided very precise masses for $\alpha$Cen A and B.
These high quality data stimulated new calibrations of the system by \citet{Thevenin02} and
 \citet{Thoul03}. The two  teams reached different results. While \citet{Thevenin02} could not
fit the seismic data without changing the masses more than 4$\sigma$ with respect to Pourbaix's data,
the second group fitted the $\alpha$Cen~A p-mode spectrum and the spectroscopic constraints
 using a single value of mixing length  parameter, and
with the new values for the masses.

Interferometric measurements with VINCI/VLTI  by \citet{Kervella03} have provided high
precision values of the angular diameter of $\alpha$Cen~A and B,
and new observations  by \citet{Carrier03} have allowed to identify p-mode frequencies
also in the B component. These new constraints have been used by \citet{Eggenberger04}. 
 Their calibration, based on a grid of models obtained by varying the
mixing-length parameters, the chemical composition, and the age, leads to a  stellar model in good agreement with
the astrometric, photometric, spectroscopic and asteroseismic data, and they assert that {\it ``the global 
parameters of the $\alpha$Cen system are now firmly constrained to an age of $t=6.52\pm0.30$~Gyr, an initial 
helium mass fraction $Y_i=0.275\pm0.010$ and an initial metallicity $(Z/X)_i=0.0434\pm0.0020$''} and that {\it 
``the mixing length parameter $\alpha$ of the B component is larger than the one of the A component''}.
These results are  quantitatively consistent with those obtained by \citet{Thoul03}, nevertheless,
both groups have performed the calibration assuming a given, but different, ``physics''.

The aim of  this work is to study, following the theoretical analysis of the utility of seismology
to constrain fundamental parameters  made by  \citet{Brown94}, the dependence of the
set of parameters obtained by fitting the observables,  on the details of the fitting procedure.
That is:
 {\it i)} the kind of constraints included in the
$\chi^2$ functional that drives the fitting procedure, as well as other effects of their uncertainties, and
{\it ii)}  the physics included in the stellar evolution theory.
Only in this way we will be able to  provide an estimation of the uncertainty in the obtained set of 
stellar parameters, and to evaluate the degree at which the present data precision 
can constrain the physics in  stellar evolution models.

To do that, and also with the prospect of dealing with a great quantity of seismic data from spatial missions 
such as
MOST \citep{Matthews98} and COROT \citep{Baglin98} and from new generation spectrograph such as HARPS 
\citep{Bouchy02a}, we have implemented
a non-linear fitting algorithm that performs a simultaneous least-square adjustment of all the observable
characteristics, the classical and  the seismic features.

The fitting method is described in Sect.~\ref{sec:cal_proc}.
In Sect.~\ref{sec:obs_constr} we
discuss the different sets of classical and seismic observables that will be included in our
$\chi^2$ quality function.
The physics included in the stellar evolutionary code is summarized in Sect.~\ref{sec:codes}.
The results of these different combinations of observables and
of different physics are presented and discussed in Sect.~\ref{sec:results}.
A special effort has been devoted to the problem of stellar convection (Sect~\ref{sec:conv}).
We will try to answer to the question about the universality of the mixing-length parameter, and to study
the effect of different convection treatments.  With respect to
the convection modelling, a controversial result was obtained by \citet{Morel00},
they reached different ages depending on whether  they used the classical MLT theory \citep{Bohm58}, or 
the FST theory by \citet[][hereafter CM91,CM92]{Canuto91,Canuto92}. 
Hence, we 
performed different calibrations changing  the  convection treatment (FST or MLT), as well as
different MLT
calibrations either with a unique or different  $\alpha$ values  for each  component.
The effects of using different equation of state, including or not gravitational settling, and adopting a
different solar mixture are discussed in
Sect.~\ref{sec:phys}.
Finally, results and conclusions are summarized in Sect.~\ref{conclu}.

\section{Calibration method}\label{sec:cal_proc}

Usually the approach to analyze stellar oscillation data is rather
conventional: {\it i}) several stellar models that bracket the
known observational constraints (typically composition, mass,
luminosity, and effective temperature) are computed ; {\it ii})
p-mode oscillation frequencies are calculated for the models; and
{\it iii}) the model oscillation spectra are compared with the
observed one.
% Comparisons between the model and observed oscillation spectra, for the most part are
%  done subjetively, with no quantitative measure of how well the spectra are matched.
 We believe that a different
 approach is needed, so that asteroseismology is directly included in the calibration procedure and
the results are not biased by a limited/subjective exploration of
the parameter space or strongly depent on an initial guess of the
model parameters.

The development and use of objective and efficient procedures to
fit stellar models to observations has
 become of evident utility in particular since seismic constraints are included in the modelling
 \citep[see e.g. ][]{Brown94}.
%does not depend on how  on having a close stellar model from which to begin.

\citet{Guenther04} have proposed a method that quantifies at which degree
the oscillation spectra as obtained  from a grid of models
parametrised in mass, age and composition reproduce the
observations. Models providing a minimum in the $\chi^2$, defined
by the differences between the theoretical and observational
frequencies, are selected for an additional inspection in a finer
grid. As noted by \citet{Guenther04} the first problem in
this kind of approach is its computational cost. Moreover the direct fit of 
 the oscillation spectra implies  a good
knowledge of the surface layers of the star, as that strongly affects
the exact frequency values. \citet{Guenther04} quantify
the uncertainty in the theoretical frequencies due to our poor
modelling of the external layers by using  the discrepancies between observed and
theoretical solar frequencies. But, how good is this estimation
for stars with different superficial gravity, chemical composition
and age? On the other hand, \citet{Eggenberger04} use the
aforementioned conventional approach, and only in a second and a
 third phase take into account
the asteroseismic  data, first the large and small frequency separations
($\Delta\nu$, $\delta\nu$), and then the frequencies.

Here we  propose a calibration method that finds the parameters of
the system by the minimization of a $\chi^2$ functional including
at the same time classical and asteroseismic observables. The
parameters of the models are, as usual, the mass, initial chemical
composition, age, parameter(s) of
 convection $\alpha$. The observables could be chosen among the  masses, T$_{\rm eff}$, $L$, $R$, [Fe/H], 
$\Delta \nu$,
$\delta \nu$  (or combinations of these frequency differences). Of
course, being $\alpha$Cen  a binary system, the same initial
chemical composition and age have to be assumed when calibrating
components A and B.

We define a quality function measuring the distance between models and observations, that is, a
goodness-of-fit measurement by:

\begin{equation}
\chi^2=\sum_{i=1}^{N_{\rm o}}{\frac{(O^{\rm obs}_i-O_i^{\rm theo})^2}{(\sigma_i^{\rm obs})^2}}
\label{eq:chisq}
\end{equation}

\noindent where $O_i^{\rm obs}$,  $\sigma_i^{\rm obs}$ and $O_i^{\rm theo}$ are respectively the observed value, observational
uncertainty  and theoretical prediction of each of the $N_{\rm o}$ (A+B components) observables considered.

 The choice of the observables included in the objective function is thoroughly described in
Section \ref{sec:obs_constr}.

In the general case of a binary system the model that generates
the observables and their derivatives has seven free parameters (or six
if $\alpha_{\rm A}=\alpha_{\rm B}$). The most substantial part of
the model consists of a stellar evolution code (CLES, see Sec.
\ref{sec:codes}) which takes as inputs the masses  of each
component ($M_{\rm A}$, $M_{\rm B}$), the initial chemical
composition of the system ($Y, Z$), the age and the convection
mixing-length parameters ($\alpha_{\rm A},\alpha_{\rm B}$), that
in general are assumed to be different for the two stars.

We compute oscillation frequencies for the models by solving the
equations of adiabatic oscillations (OSC), and  determine $\langle\Delta \nu\rangle$
and $\langle\delta \nu\rangle$ for the degrees $\ell =0,1,2,3$.
This is not done by 
 a least square fit to the computed frequencies, based on
the asymptotic properties of low degree modes, but by making the
average of the theoretical  separations in the domain of observed
radial order $n$  ($n=15-25$, for $\alpha$Cen~A, and $n=17-27$ for
$\alpha$Cen~B). The observed average separations have been
determined from observed  frequencies in the same way.

  For each component the evolution code provides the stellar luminosity, the radius and the effective
  temperature, as well as the model quantities required in the subsequent calculations of the oscillation
  frequencies. In the  calibration process, the derivatives of the observable quantities
  are obtained varying each of the parameters ($M_{\rm A}$,$M_{\rm B}$, Z, Y, $\alpha_{\rm A}$,
   $\alpha_{\rm B}$, and the age).
We do not derive colors or visual magnitudes, and  we do not
include orbital elements
  such as apparent semi-major axis ($a'$), orbital period ($P_{\rm orb}$) or parallax. We assume
  the parallax as determined by \citet{Soderhjelm99} and the values of observables based on it.
  The masses, however,  are considered as parameters and also as observable
  quantities in our calibrations.

In the calibration of a binary
 system the large number of variables involved, both in terms of model parameters and
 observables, suggests the use of a least-squares based fitting procedure. This  is
 particularly useful in this kind of calibration as we fit
at the same time classical and seismological observables without
making first a selection based on the HR location of the system.

   As shown in \cite{Brown94} the observable quantities depend on the parameters in a complex
    way: stellar evolution
   is not   a  linear problem. Most of the observables
are influenced by several parameters, and hence the connection
between observables and parameters that could conceptually seem  to us the 
simplest one will not always provide the correct results.

\subsection{Optimization algorithm}\label{sec:algo}

We use the gradient-expansion algorithm known as Levenberg-Marquardt method.
This algorithm combines the advantages of an expansion method, i.e.  rapid convergence
 close to the minima,
with those of the gradient-search, that is, a rapid approach  to
a far away minimum. This method has as well the strong
advantage of being reasonably insensitive to the starting values
of the parameters.

At each step the fitting function is linearized calculating
numerical derivatives (centered differences) of the observables
with respect to each model parameter. The displacement in the
parameter space, leading to a lower value of $\chi^2$, is
calculated following the prescription of \citet{Bevington}, p.
161-164 and the iterative procedure is ended when $\chi^2$ no
longer changes more than 2\%.
In calibrations with  $M_{\rm A}$,$M_{\rm B}$, Z, Y, $\alpha_{\rm A}$,
   $\alpha_{\rm B}$, and the age as  model parameters, convergence is typically achieved in 3-4 iterations and at each of them 
 the computation of 16 evolutionary tracks is needed to evaluate centered derivatives.
The result of such a local minimization could be sensitive to the
initial guess of the parameters; therefore, in order to get a more reliable 
final solution we  perform several runs starting from 
different  points in the parameter space. The effect we
find is limited to a variation of the number of iterations needed
to reach the minimum $\chi^2$, whereas the final parameters of the
system differ much less than their uncertainty.
 On the other hand, building
a 6-dimensional dense grid of models seems impractical considering
the aim  of evaluating the effects
of using different physical prescriptions in our models (e.g.
different equation of state, different metal mixture etc).

  It is sometimes nonetheless possible to make fairly direct connections between the observables and the
  parameters,  particularly when one observable is much better determined than the rest.
  The solution is determined by the parameter that is known with very small
  uncertainty.
The uncertainties in the parameters for these fits are calculated
from the diagonal terms in the error matrix (inverse of curvature
matrix in the parameter space) and are, in general considerably
larger than the uncertainties obtained in the grid- and
gradient-search methods. The latter are obtained by finding the
change in each parameter to produce as change of $\chi^2$ of 1
from the minimum values, without re-optimizing the fit, while there
is a strong suggestion that correlations among the parameters play
an important role in fitting \citep[see e.g.][]{Bevington}.

\section{The choice of the observational constraints}\label{sec:obs_constr}
\subsection{Non-asteroseismic constraints}

Due to the proximity of the $\alpha$Cen binary system, the
precision on the measurement of its trigonometric parallax is
potentially very high. Unfortunately, some discrepancies have
appeared among the most recent published values 
\citep[see Table 10 in][]{Kervella03}. \citet{Guenther00}  studied
the uncertainty in the stellar
 parameters due to the different parallax values. This, indeed  affects the determination of  mass, 
 luminosity and  radius.  Following  \citet{Eggenberger04} we adopt $\pi=747.1\pm 1.2$~mas \citep{Soderhjelm99}
and, therefore, the corresponding mass values determined by \citet{Pourbaix02}: ($M_{\rm 
A}=1.105\pm0.007$~\msol, $M_{\rm
B}=0.934\pm0.006$~\msol)
 and the  radii : ($R_{\rm A}=1.224\pm0.003$~\rsol; $R_{\rm
B}=0.863\pm0.005$~\rsol)
\citep{Kervella03}.
% both based on that value of parallax.

As in the case of  the parallax, there is a large scatter in the
published values of other quantities, such as \teff, luminosity,
and metallicity. We decided to use the same values adopted by
\citet{Eggenberger04} in order to  have a  reference
model. \citet{Eggenberger04} took as \teff\ for the component
A a value to encompass those given by two spectroscopic
determinations, the one from \citet{Neuforge-Verheecke97}
(\teff$_{\rm A}$=5830$\pm 30$~K,
 \teff$_{\rm B}$=5255$\pm 50$~K), and the one
by \citet{Morel00} based on  a re-analysis of \citet{Chmielewski92} 
spectra (\teff$_{\rm A}$=5790$\pm 30$~K, \teff$_{\rm
B}$=5260$\pm 50$~K), used  respectively in the $\alpha$Cen
calibrations by \citet{Thoul03} and \citet{Thevenin02}.
We will use the effective temperature as constraint in our
minimization method only in one of the calibrations (A1t,B1t, in
Table~\ref{tablepar}), since the precise determination of the
radius provides a narrower
domain in the space of observable quantities.
The luminosity values adopted by \citet{Eggenberger04} come
from a new and weighted calibration of previous Geneva photometric
data,  where they have also coherently taken into account  the
effective temperatures and the parallax. These values cover the
domain considered by \citet{Thevenin02} that considered an
error bar twice smaller, and a lower luminosity for the component
B. The luminosity values,  directly determined  from the
adopted radius and effective temperature, are in very good
agreement with the those determined by \citet{Eggenberger04}:
$L_{\rm A}/L_{\odot}=1.518 \pm 0.06$, $L_{\rm B}/L_{\odot}=0.507
\pm 0.025$. On the other hand, the values determined by \citet{Pijpers03}
 for $L_{\rm A}$ are much larger and only marginally
overlap the values considered here.

The precise values of masses and radius provide also precise
values of the surface gravity for both stars: $\log g_{\rm
A}$=$4.305\pm 0.005$ and $\log g_{\rm B}$=$4.536 \pm 0.008$, while the
spectroscopic values determined by \citet{Neuforge-Verheecke97} 
are $\log g_{\rm A}=4.34\pm 0.05$ and $\log g_{\rm B}=4.51
\pm 0.08$. These values were used by \citet{Thoul03} to fix
the  luminosity domain , leading to  higher central values and
to larger error bars with respect to those determined by
\citet{Eggenberger04} and \citet{Thevenin02}.

Also for the metallicity of both components there is no complete
agreement in the literature: [Fe/H]$_{\rm A}=0.20 \pm 0.02$, and
[Fe/H]$_{\rm B}=0.23 \pm 0.03$ from \citet{Morel00}, and
[Fe/H]$_{\rm A}=0.25 \pm 0.02$, and [Fe/H]$_{\rm B}=0.24 \pm 0.03$
from \citet{Neuforge-Verheecke97}. The uncertainty in the
observable $Z/X$ is quite large, if we take into account also the
10\% in the $(Z/X)_{\odot}$. We have taken
the value adopted by \citet{Thoul03}, that is $Z/X=0.039 \pm
0.006$, the same for both stars. The detailed abundance analysis
of $\alpha$ Cen A and B carried out in \citet{Neuforge-Verheecke97} suggested
no evidence for a different metal mixture relative to the sun,
therefore all our models were computed assuming the solar mixture
by \citet{Grevesse93}, except for the calibration (A5,B5) in
which we have considered the recently determined solar metal
abundances \citep{Asplund04,Asplund05} that implies $(Z/X)_{\odot}=0.0177$.

The non-asteroseismic constraints used in this work are summarized in Table \ref{tab:nonastero}.
\begin{table}
\caption{\small Non-asteroseismic constraints. References(1):\cite{Eggenberger04}; (2): \cite{Thoul03}.}
\label{tab:nonastero}
\begin{tabular}{ccccc}
  \hline
  & A & B & Ref \\
\hline
\hline
M/\msol & 1.105$\pm$0.007& 0.934$\pm$0.006& (1) \\
\teff & 5810$\pm$50 & 5260$\pm$50 & (1)\\
R/\rsol & 1.224 $\pm$ 0.003 & 0.863$\pm$0.005& (1) \\
L/\lsol & 1.522$\pm$0.030 & 0.503$\pm$0.020&(1) \\
Z/X & 0.039$\pm$0.006 &0.039$\pm$0.006& (2)\\
 \hline
 \end{tabular}
\end{table}

\subsection{Seismic constraints}\label{sec:seis_constr}

Solar-like oscillations generate periodic motions of the stellar
surface with periods in the range of 3-30~min
 and with extremely small amplitudes. Frequency and amplitude of each oscillation mode depend on the physical
 condition prevailing in the layers crossed by the waves and provide a powerful seismological tool.
 Helioseismology led to major improvements in the knowledge of Sun structure and to revision of the ``standard 
solar
 model". The potential utility of seismology applied  to other stars, in particular $\alpha$Cen,
 to constrain the stellar parameters was extensively studied in \citet{Brown94},
 and also by \citet{Guenther00}.

 Several groups had made thorough attempts  to detect the signature of p-mode oscillations in $\alpha$Cen A, but 
their
 results were not confirmed. Only recently \citet{Bouchy02}, from high precision radial velocity
 measurements with the CORALIE echelle spectrograph  have yielded a clear detection of p-mode oscillation, and
 identified several modes between 1800 and 2900 $\mu$Hz , and with an envelope amplitude of about
 $31 ~{\rm cm}\,{\rm s}^{-1}$.
 %The seismological parameters deduced from their observations were in full agreement
 %with the expected values scaling from the Sun (Kieldsen \& Bedding 1995)
 %giving the frequency of the greatest mode $\nu_{\rm max}=2300\mu{\rm Hz}$, a large spacing 
 %$\Delta\nu_{0}=105.8$,
 %and the oscillation peak amplitude $A_{\rm osc}=31.1 {\rm cm}{rm s}^{-1}$.
Assuming that frequency modes $\nu_{n \ell}$ satisfy the
simplified asymptotic relation \citep{Tassoul80}:
\begin{equation}
\nu_{\ell n} \approx \Delta\nu_0\left(n+\frac{\ell}{2}+\epsilon\right)-\ell(\ell+1)\frac{\delta\nu_{02}}{6}
\end{equation}
\noindent and assuming the parameter $\epsilon$ near the solar one
(1.5) they estimated: $\Delta\nu_0=105.5\pm 0.1\, \mu{\rm Hz}$,
$\delta\nu_{02}=5.6 \pm 0.7\,\mu{\rm Hz}$, $\epsilon=1.40 \pm 0.02$
and identified 28 p-modes with degree $\ell=0,1,2$ and order
between $n=15$ and $n=25$.

Notice that the given errors  come from the autocorrelation
algorithm, but we must keep in mind that their frequency
resolution is only 0.93~$\mu{\rm Hz}$, and that they derive an
uncertainty in the frequency determination equal to 0.46~$\mu{\rm
Hz}$. They also point out that an error of $\pm 1.3 \mu{\rm Hz}$
could have been introduced at some identified mode frequency, that
could explain the dispersion of mode frequency around the
asymptotic relation.  In particular higher observational uncertainty could affect mainly the $\ell=2$ modes that
determine the value of $\delta\nu_{02}$ for the lower and higher
frequency $\delta\nu_{02}$(n=16 and 25).

 \citet{Carrier03} have also detected solar-like oscillations in the fainter component, $\alpha$Cen~B. 
 Only twelve  frequencies,
between 3000 and 4600~$\mu{\rm Hz}$ have been kept in the final
list of identified p-modes, four of them  with a detection  level
lower than 3$\sigma$, and it is 
recommended to take them with caution. As for component A the
frequency resolution is 0.93~$\mu{\rm Hz}$. The large and small
separations, determined by autocorrelation of the asymptotic
relation, are respectively $\Delta\nu_0=161.1\pm 0.1$~$\mu{\rm Hz}$
and $\delta\nu_{02}=8.7 \pm 0.8$~$\mu{\rm Hz}$. We must note here
that the value derived for $\delta\nu_{02}$ comes from only few
p-modes. In fact, from their frequency table is only possible to
obtain two values: $\delta\nu_{02}(n=21)=10.0 \mu{\rm Hz}$ and
$\delta\nu_{02}(n=23)=7.0 \mu{\rm Hz}$. \citet{Carrier03}
expect a rotational splitting $\sim 0.3 \mu{\rm Hz}$ that, given
the frequency resolution, could imply an increase of the
uncertainty of frequencies for modes of degree $\ell=1$ and
$\ell=2$.

 Very recently, new observations of this system by \citet{Kjeldsen04} have confirmed the values determined
 by \citet{Bouchy02} concerning the frequency separations of component A. However, component B with
 this new more precise data shows
 $\Delta\nu_0=161.4 \mu{\rm Hz}$ and $\delta\nu_{02}=10.1  \mu{\rm Hz}$.
 Our computations have been done before these values were available, therefore, we will not take them into
 account in our calibration. Notice that, nevertheless, those values are anyway reached and imposed
 by the other observables used in some of our calibrations (A3,B3), see Sec. \ref{sec:results}.

 How should these seismological observations be used to constrain our stellar models?
The classical way is to use the large and small separations to characterize
the power spectrum of solar-like oscillations.
The standard asymptotic theory of stellar oscillations \citep{Tassoul80} relates the averages
values of high-radial order / low-degree small and large separations to conditions
 in the stellar core ($\delta\nu$)  and to the mean density of the star ($\Delta\nu$).
 Recently \citet{Guenther04} and  \citet{Metcalfe05} have proposed to use directly the
 p-mode frequencies as observables to constrain the stellar models.

\citet{Brown94} theoretically analyzed the case in which the
individual frequencies are included  as observables. Their purpose was to
illustrate the potential loss of information resulting from
representing the spectra  in terms of the  large and  small separations 
derived from the expected asymptotic
 behavior. The dominant  source of frequency changes is very close to the  stellar surface.
 It could be difficult to disentangle these effects from the uncertainties in the treatment
  of the physics of the outer layers,
where non-adiabaticity  and dynamics effects of convection have to be taken into account.

 The oscillation frequencies, the large and small separations depend on the structure of both the inner and the 
outer
 layers of a star, so model fitting and testing techniques to probe the interior structure
 of the stars are dependent on our having a good understanding of the structure of the outer
 layers. But these are just the layers where our ignorance is greatest; non-adiabatic convection
 is important but not understood, the oscillations are non-adiabatic in the surface
 layers and the structure of real stellar atmosphere is poorly understood.
  For example the oscillation frequencies predicted by the reference solar models (S96)\citep{JCD96} differ from
  the observed values up to 10$\mu{\rm Hz}$ at the higher end of the observed frequency range.

In a first step, we will include as seismic constraints in our fitting algorithm
the  combinations of frequencies: the large
$$\Delta\nu_{n,\ell}=\nu_{n,\ell}-\nu_{n-1,\ell}$$ and small
$$\delta\nu_{n,\ell}=\nu_{n,\ell}-\nu_{n-1,\ell+2}$$ frequency separations
defined from the identified p-mode frequencies for both components.

\begin{figure}
\resizebox{\hsize}{!}{\includegraphics[]{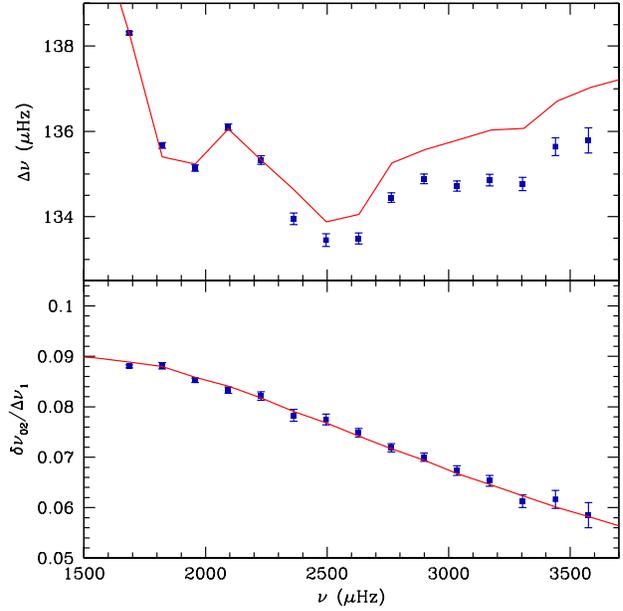}}
\caption{\small Solar large frequency difference $\Delta\nu_{n,1}$ (upper panel) from standard seismic solar 
model S96 \citep{JCD96}
(solid line), compared to the observational solar large separation (dots) \citep{Basuetal97}.
Lower panel: as upper panel but for the ratio $r_{02}$.}
\label{fig:sunls}
\end{figure}

As discussed in \citet{JCD95} and \citet{DiMauro03}, care has to be taken when considering as a constraint in
the modelling the large separation, as its averaged value at high frequencies could be influenced by
near-surface effects as well. This is in fact the case when comparing the observed and the  predicted
low-degree large separations of the Sun (see Fig. \ref{fig:sunls}), where the disagreement of the order of
 a $\mu$Hz is related to a simplified treatment of the model outer layers.

With the aim of checking whether the calibration we perform considering $\langle\Delta\nu\rangle$ and 
$\langle\delta\nu\rangle$ is
not biased by a simplified treatment of the outer structure in our models, we considered the effect on
the calibration of choosing as seismic constraint $r_{02}$, the ratio between the small and large frequency
separations defined by:
\begin{equation}
r_{02}(n)=\frac{\delta\nu_{n,0}}{\Delta\nu_{n,1}}=\frac{\nu_{n,0}-\nu_{n-1,2}}{\nu_{n,1}-\nu_{n-1,1}}
\label{eq:rox}
\end{equation}

\noindent for 6  different orders $n$ of the
component A. This combination of frequencies, as presented in
\citet{Roxburgh03}, is to a great accuracy independent of the outer
layers of the star, and therefore represents a reliable indicator
of the conditions in the deep stellar interior.

\section{Stellar models}\label{sec:codes}
All stellar model sequences are calculated using the  CLES code (Code Li\'egeois d'Evolution  Stellaire).
 The opacity tables are those of OPAL96 \citep{Iglesias96} complemented at $T < 6000$~K with \citet{Alexander94}
 opacities.
The  relative mixture of heavier than helium elements, used in the opacity and equation of state tables, is
the solar one according to  \citet{Grevesse93}.
The nuclear energy generation routines are based on the cross sections by \citet{Caughlan88} and screening
factors
 from \citet{Salpeter54}.
CLES allows the choice between two equations of state: CEFF
\citep{JCD92} and OPAL01 \citep{Rogers02}. Most of the models
have been computed using OPAL01, but we have also made a
calibration using CEFF to study the capability of seismology to
constrain the EoS.

All our stellar models were obtained from  evolutionary tracks including  the pre-main sequence
  phase, and ending  at $\sim$9~Gyr.
The stellar models have approximately  1200 shells,
the last one corresponding to T=T$_{\rm eff}$ as determined using
as boundary conditions those given by the \citet{Kurucz98}
atmosphere models. Furthermore, for the computation of
oscillations we have added atmospheric layers from $T=T_{\rm
eff}$ up to $\tau=10^{-4}$.

We have computed models where the convection is treated both using
the Mixing Length Theory \citep[MLT][]{Bohm58} with the formalism
described in  \citet{Cox}, and  the Full Spectrum of Turbulence
\citep[FST][]{Canuto96a} with a formalism similar to the one
used by \citet{Morel00} or \citet{Bernkopf98}. That means 
convective fluxes as given by \citet{Canuto96a}, but  another
prescription of the scale length. The parameter  $\alpha$ of the
MLT and the corresponding in the FST are considered as free
parameters of the model calibration, and the values obtained are
compared with the values required in the solar calibration for the
same physics.

Finally we have computed stellar models with and without 
gravitational settling  of helium and metals. The microscopic
diffusion formulation is that given by  \citet{Thoul94}, solving
the \cite{Burgers69} equations for H,
 He and Z  and thus considering diffusion due both to thermal and concentration gradients. We also assume complete ionization and the effect of radiative
acceleration is ignored.
This assumption is completely justified for the precision of the models and for the masses considered
\citep{Turcotte98}.

 \begin{figure}[t]
\resizebox{\hsize}{!}{\includegraphics[angle=-90]{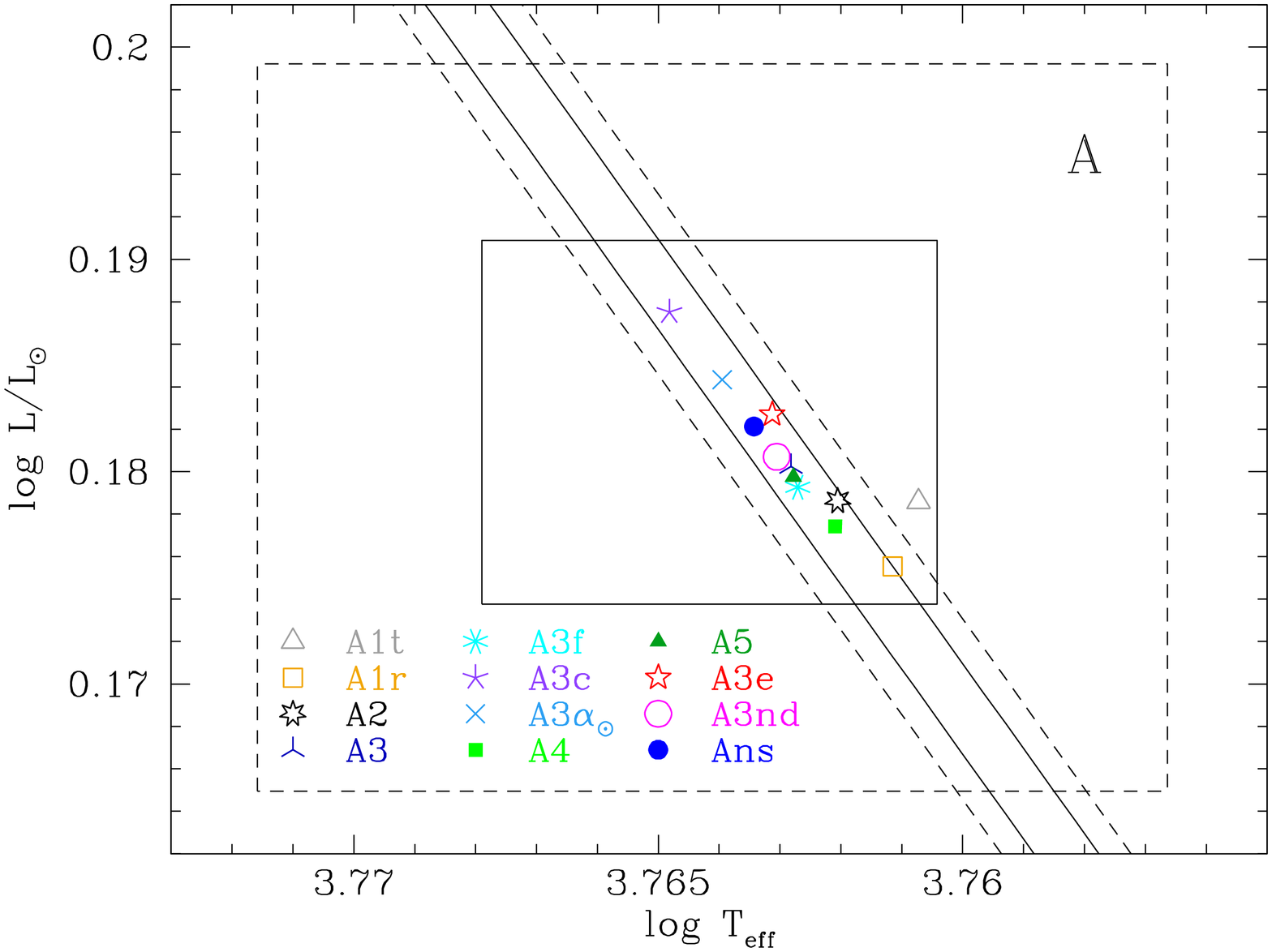}}
%\end{figure}
%\begin{figure}[t]
\resizebox{\hsize}{!}{\includegraphics[angle=-90]{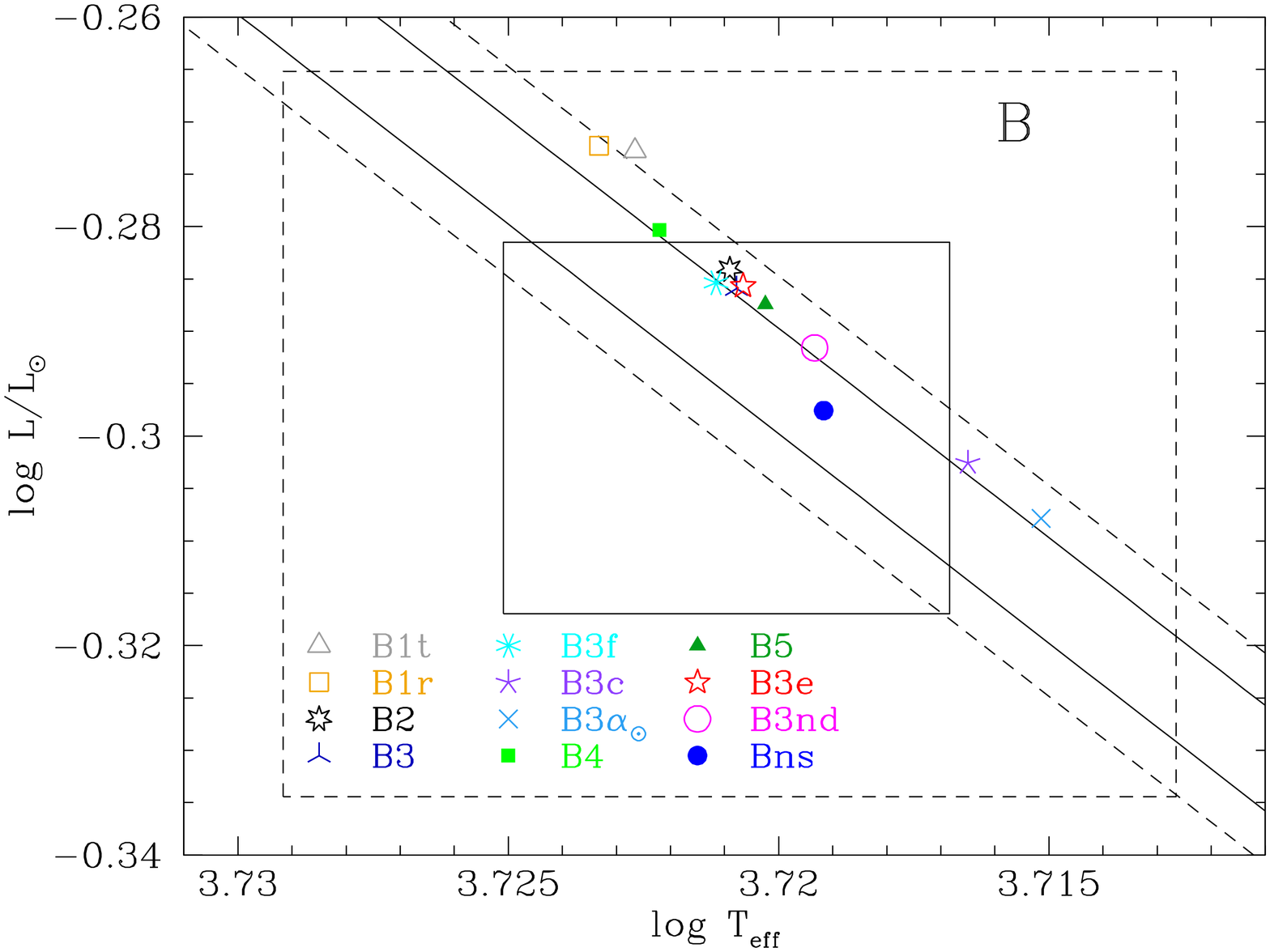}}
\caption{\small HR diagram location of models with stellar parameters from
differing fits. Labels corresponds to those in Tables~\ref{tablepar}, and ~\ref{tableobs}.
Upper and lower panel correspond respectively to components A and B. The error boxes for $T_{\rm eff}$,
$\log L/L_{\odot}$, and radii correspond to 1$\sigma$ (solid line) and $2\sigma$ (dashed-line)}
\label{fig:hr}
\end{figure}

\section{Results}
\label{sec:results}
 In Table~\ref{tablepar} we list all the
fits  of the binary system $\alpha$Cen. The calibrations
differ in the choice of the observables included in the
objective function ($\chi^2$) and in the used physics. The first
column in Table~\ref{tablepar} identifies the models, the second,
third and fourth   columns describe the physics; the fifth one resumes
the characteristic of the seismic and non--seismic constraints used
in the calibration process, while all the following columns give the
values of the parameters (and the uncertainty in each of them)
providing in each case the minimum  $\chi^2$.
 The values of the observable quantities derived for each set of parameters is listed in
Table~\ref{tableobs}. There, the last column $\chi^2_{\rm R}$ is
the the value of the ``reduced'' $\chi^2$, defined
 as:
\begin{equation}
\chi^2_{\rm R}=\frac{\chi^2}{N_{\rm o}-N_{\rm p}} \label{eq:redchisq}
\end{equation}
were $N_o$ is the number of observables and $N_p$ the number of
model parameters.

In the three following subsections we shall analyze the effect of
changing the non-seismic constraints, the seismic constraints and
finally the physics used in the models, that is, a different
treatment of convection, different equation of state, different
solar  mixture, models  including or
not gravitational settling, and the effect of considering
overshooting in our models.

Fig.~\ref{fig:hr} shows the HR location of each of these fitted models. We have indicated the
error boxes in \teff\, $\log  L/L_{\odot}$, and radius, corresponding to 1$\sigma$ (solid line), and
2$\sigma$ (dashed-line).

\begin{figure}
\resizebox{\hsize}{!}{\includegraphics[angle=-90]{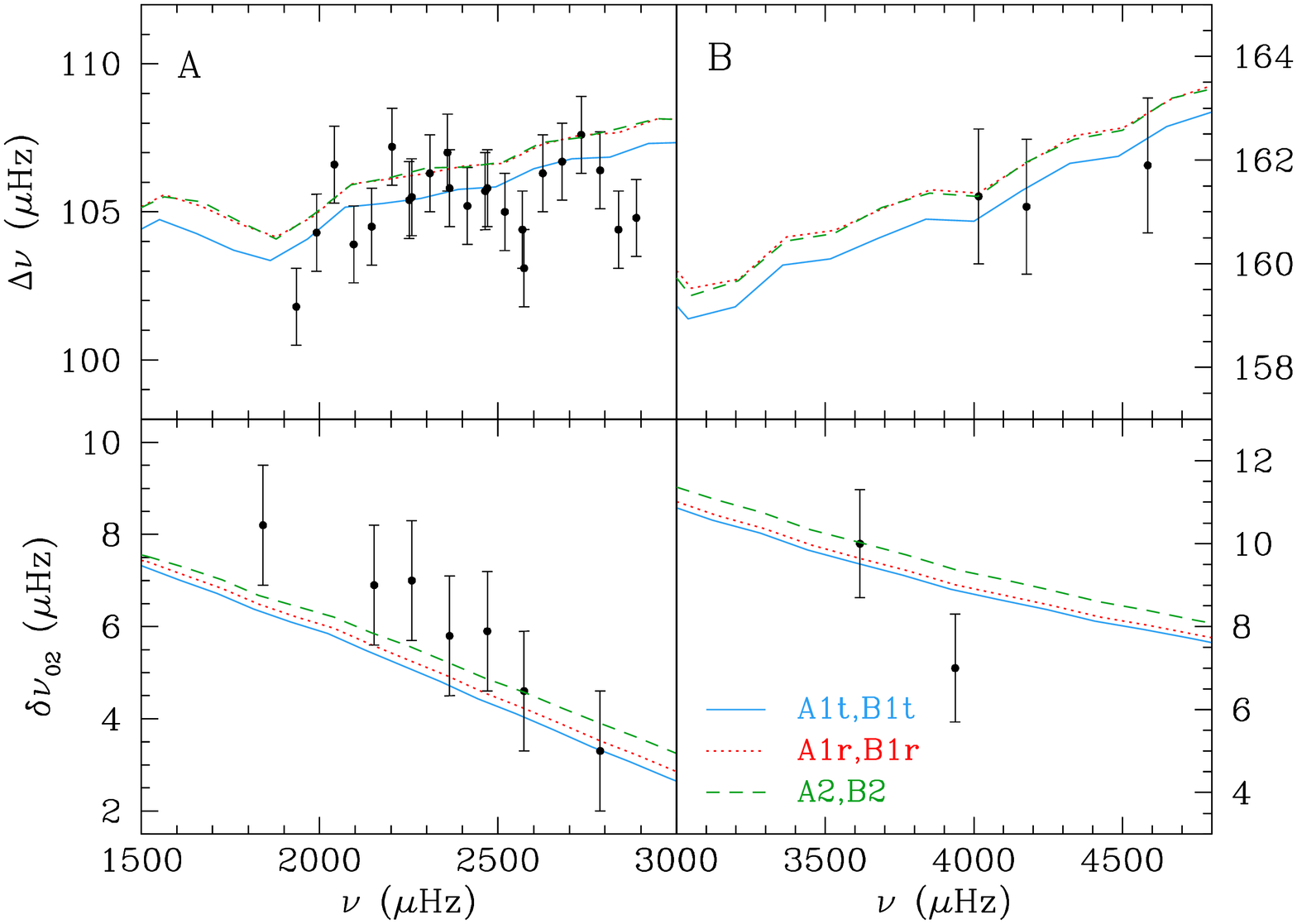}}
\caption{\small Large (upper panels) and small (lower panels) separations for the A (left) and B (right) 
components
of $\alpha$Cen system. Points represent the observational values with their error bars assuming
an error in frequencies equal to $2\sigma$. All these curves
correspond to calibrations performed including in the $\chi^2$ functional the average of the observational large 
and
small separations, but different classical observational constraints: masses fixed to the value given by 
\citet{Pourbaix02} and  \teff\ (solid lines); radii (dotted-lines); and masses variable as parameter and  radii
 (dashed-lines). All the curves were computed from stellar models including gravitational settling,
and MLT treatment of convection with two mixing length adjustable parameters.  For clarity only $\ell=1$ theoretical large separations are shown in the upper panels.}\label{fig:seismo1}
\end{figure}

\begin{table*}
\caption{\small Sets of parameters of fitted models. Meaning of numeric labels: 
1 (fixed masses and $\Delta\nu$, $\delta\nu$ as seismic constraints); 
2 (as 1 but with variable masses); 3 (as 2 but using $r_{02}(n)$ as seismic constraint);
4 (as 3 but including convective overshooting) and 
5 (as 3 but using Asplund et al. (2005) instead of Grevese \& Noels (1993)).
Meaning of  alphabetic labels: nd (non diffusion models); f (FST convection treatment); 
e (CEFF EoS instead of OPAL01 one);
c (a unique mixing-length parameter);
ns (fit without seismic constraints);  only for the case 1, t  refers to effective temperature as constraint, and
r  to radius as constraint.}

\begin{tabular}{c|ccc|c|cccccccccc}
%\hline
Model  &  Conv  &  Diff  &  EoS  & fitting &  M/\msun  & $\epsilon$ & $\alpha$  & $\epsilon$ &  Age & $\epsilon$ 
&  $Y_0$ & $\epsilon$ &  $Z_0$ & $\epsilon$ \\
% mfix averaged_sep teff
\hline
\hline
A1t     & MLT&  Y  & OPAL &$ T_{\rm eff}$, $\rm M_{\rm fix}$ & 1.105   &  & 1.99 & 0.11& 7.0  & 0.5 &  0.273 & 
0.016 & 0.0316  & 0.003 \\
B1t     & MLT&  Y  & OPAL & $\langle\Delta\nu\rangle, \langle\delta\nu\rangle$ & 0.934   & & 2.22 & 0.07& 7.0  & 
0.5 &  0.273 & 0.016 & 0.0316  & 0.003 \\
% mfix averaged_sep radii
\hline
A1r     & MLT&  Y  & OPAL & $M_{\rm fix}$& 1.105  & & 2.00  & 0.11 & 6.8  & 0.5 & 0.276  & 0.015 & 0.0325  & 
0.003\\
B1r     & MLT&  Y  & OPAL & $\langle\Delta\nu\rangle, \langle\delta\nu\rangle$& 0.934  & & 2.24  & 0.07 & 6.8  & 
0.5 & 0.276  & 0.015 & 0.0325  & 0.003\\
% mfix averaged_sep radii+atmo
%A1ra    & MLT&  Y  & OPAL & $\rm  R$, $\rm M_{\rm fix}$, & 1.105  & & 2.00  & 0.11 & 6.8  & 0.6 & 0.275  & 
%0.016 & 0.0321 & 0.003 \\
%B1ra     & MLT&  Y  & OPAL & Atmo& 0.934  & & 2.27  & 0.07& 6.8  & 0.6 & 0.275  & 0.016 & 0.0321 & 0.003 \\
% \hline
% mfix averaged_sep no diff
\hline
A1nd     & MLT&  N  & OPAL & $M_{\rm fix}$ & 1.105  &  &   1.84  & 0.10 &   7.1  & 0.6 &   0.265  & 0.015 &   
0.0284   & 0.003\\
B1nd     & MLT&  N  & OPAL & $\langle\Delta\nu\rangle, \langle\delta\nu\rangle$& 0.934  &  &   1.99  & 0.05 &   
7.1  & 0.6 &   0.265  & 0.015 &   0.0284   & 0.003\\
% mfix averaged_sep CEFF
\hline
A1e     & MLT&  Y  & CEFF  & $M_{\rm fix}$ & 1.105  &  &  1.93   & 0.10 &   6.7  & 0.6 &   0.281  & 0.015 &   
0.0324   & 0.003\\
B1e     & MLT&  Y  & CEFF  &$\langle\Delta\nu\rangle, \langle\delta\nu\rangle$ & 0.934  &  &  2.08   & 0.06 &   
6.7  & 0.6 &   0.281  & 0.015 &   0.0324   & 0.003\\
\hline
% mfix rox + atmo
A1-3    & MLT &  Y  & OPAL &  $M_{\rm fix}$, & 1.105  & & 1.86  & 0.06 & 6.0  & 0.4 & 0.277  & 0.017 & 0.0302 & 
0.003 \\
B1-3    & MLT&  Y  & OPAL & $r_{02}(n)$ & 0.934  & & 2.18  & 0.05& 6.0  & 0.4 & 0.277  & 0.017 & 0.0302 & 0.003 
\\
\hline
% changing mass + averaged_sep
A2     & MLT&  Y  & OPAL & $M_{\rm var}$& 1.104  & 0.006 & 1.96  & 0.11 & 6.4  & 0.6& 0.282  & 0.016& 0.0326  & 
0.003\\
B2     & MLT&  Y  & OPAL & $\langle\Delta\nu\rangle, \langle\delta\nu\rangle$& 0.926  & 0.005 & 2.11  & 0.10 & 
6.4  & 0.6& 0.282  & 0.016& 0.0326  & 0.003\\
\hline
% changing mass + averaged_sep + 1 MLT
A2c     & MLT&  Y  & OPAL & $M_{\rm var}$ $\alpha_{\rm A}=\alpha_{\rm B}$& 1.107  & 0.005 & 1.78  & 0.08 & 5.4  
& 0.5& 0.284  & 0.016& 0.0305  & 0.003\\
B2c     & MLT&  Y  & OPAL & $\langle\Delta\nu\rangle, \langle\delta\nu\rangle$& 0.919  & 0.004             &  ``   
& ``   & 5.4  & 0.5& 0.284  & 0.016& 0.0305  & 0.003\\
\hline
% changing mass + averaged_sep + FST
A2f     & FST&  Y  & OPAL &$M_{\rm var}$ & 1.105 & 0.006   & 0.77  & 0.05 & 6.4  & 0.6 & 0.282  & 0.016 & 0.0325  
& 0.003\\
B2f     & FST&  Y  & OPAL &$\langle\Delta\nu\rangle, \langle\delta\nu\rangle$ & 0.927 & 0.005   & 0.89  & 0.05 & 
6.4  & 0.6 & 0.282  & 0.016 & 0.0325  & 0.003\\
\hline
% changing mass + averaged_sep + 1 punto SSEP B
%A2s     & MLT&  Y  & OPAL &$\rm M_{\rm var}$ & 1.105 & 0.006   & 1.92  & 0.11 & 6.1  & 0.6 & 0.285  & 0.016 & 
%0.0325  & 0.003\\
%B2s     & MLT&  Y  & OPAL & 1 Point $\delta\nu_B$& 0.926 & 0.005   & 2.07  & 0.10 & 6.1  & 0.6 & 0.285  & 0.016 
%& 0.0325  & 0.003\\
%\hline
% rox mlt
A3     & MLT&  Y  & OPAL & $M_{\rm var}$ & 1.110  & 0.006 &   1.90  & 0.07  &   5.8  & 0.2 &   0.284  & 0.014 &   
0.0322   & 0.003\\
B3     & MLT&  Y  & OPAL & $r_{02}(n)$ & 0.927  & 0.005 &   2.05  & 0.08 &   5.8  & 0.2 &   0.284  & 0.014 &   
0.0322   & 0.003\\
\hline
% rox 1 param mlt
A3c     & MLT&  Y  & OPAL &$M_{\rm var}$, $\alpha_{\rm A}=\alpha_{\rm B}$& 1.113  & 0.006 &   1.95  & 0.05 &   
5.8  & 0.5 &   0.276  & 0.017 &   0.0317   & 0.003\\
B3c     & MLT&  Y  & OPAL &$r_{02}(n)$ & 0.922  & 0.004 &    ``  & `` &   5.8  & 0.5 &   0.276  & 0.017 &   
0.0317   & 0.003\\
 \hline
% rox mlt sun
$\rm A3\alpha_{\odot}$     & MLT&  Y  & OPAL &$M_{\rm var}$, $\alpha=\alpha_{\odot}$ & 1.114  & 0.006 &   1.91  &  
&   5.7  & 0.2 &   0.283  & 0.010 &   0.0314   & 0.002\\
$\rm B3\alpha_{\odot}$    & MLT&  Y  & OPAL & $r_{02}(n)$ & 0.921  & 0.004 &    ``  &  &   5.7  & 0.2 &   0.283  
& 0.010 &   0.0314   & 0.002\\
\hline
% rox fst atmos no bug
A3f     &  FST  &  Y  & OPAL &$M_{\rm var}$ & 1.110  & 0.006 &   0.74  & 0.03  &   5.8  & 0.2 &   0.285 & 0.015 
&   0.0323   & 0.003 \\
B3f     &  FST  &  Y  & OPAL & $r_{02}(n)$ & 0.927  & 0.005 &   0.86  & 0.03 &   5.8  & 0.2 &   0.285  & 0.015 &   
0.0323   & 0.003\\
\hline
% rox no diff
A3nd     & MLT&  N  & OPAL &$M_{\rm var}$& 1.108  & 0.006 &   1.84  & 0.07 &   6.3  & 0.4 &   0.270  & 0.015 &   
0.0281   & 0.003\\
B3nd     & MLT&  N  & OPAL & $r_{02}(n)$ & 0.929  & 0.005 &   1.99  & 0.07 &   6.3  & 0.4 &   0.270  & 0.015 &   
0.0281   & 0.003\\
% rox CEFF
\hline
A3e     & MLT&  Y  & CEFF &$M_{\rm var}$ & 1.109  & 0.006 &   1.85  & 0.07 &   5.7  & 0.3 &   0.288  & 0.015 &   
0.0322   & 0.003\\
B3e     & MLT&  Y  & CEFF & $r_{02}(n)$& 0.927  & 0.005 &   1.93  & 0.07 &   5.7 & 0.3 &   0.288  & 0.015 &   
0.0322   & 0.003\\
\hline
% rox fst
%A4     &  FST  &  Y  & OPAL &$\rm M_{\rm var}$, Rox & 1.110  & 0.006 &   0.74  & 0.03  &   5.8  & 0.4 &   0.284  
%& 0.015 &   0.0321   & 0.003 \\
%B4     &  FST  &  Y  & OPAL & & 0.927  & 0.005 &   0.87  & 0.03 &   5.8  & 0.4 &   0.284  & 0.015 &   0.0321   
%& 0.003\\
% \hline

% rox fst atmos no bug + 2 sigma massB
%A4as     &  FST  &  Y  & OPAL &$\rm M_{\rm var}$, Rox & 1.107  & 0.006 &   0.73  & 0.03  &   5.7  & 0.4 &   
%0.288 & 0.015 &   0.0323   & 0.003 \\
%B4as     &  FST  &  Y  & OPAL &  $2\sigma  M_B$ & 0.921  & 0.007 &   0.83  & 0.04 &   5.7  & 0.4 &   0.288  & 
%0.015 &   0.0323   & 0.003\\
% rox overshoot
A4  &  OV  &  Y  & OPAL &$M_{\rm var}$ & 1.112  & 0.006 &  1.77   & 0.05  &   5.2  & 0.2 &   0.285  & 0.015 &   
0.0299   & 0.003 \\
B4  &  OV  &  Y  & OPAL & $r_{02}(n)$& 0.925  & 0.005 &  1.98   & 0.07 &   5.2  & 0.2 &   0.285  & 0.015 &   
0.0299   & 0.003\\
% rox mix2
\hline
A5 & MLT&  Y  & OPAL &$M_{\rm var}$, A04 mix & 1.109  & 0.006  &  1.74   & 0.06  &   5.9  & 0.3 &   0.280  & 
0.013  &   0.0239   & 0.002\\
B5  & MLT&  Y  & OPAL & $r_{02}(n)$ & 0.927  & 0.005 &  1.84   & 0.07 &   5.9  & 0.3 &   0.280  & 0.013 &   
0.0239   & 0.002\\
% no seismology
\hline
Ans     & MLT&  Y  & OPAL &$M_{\rm var}$,& 1.105  & 0.007 &   2.40  & 0.33  &   8.9  & 1.8 &   0.259  & 0.021 &   
0.0340   & 0.003 \\
Bns     & MLT&  Y  & OPAL & No Seismo  & 0.934  & 0.006 &   2.61  & 0.31 &   8.9  & 1.8 &   0.259  & 0.021 &   
0.0340   & 0.003\\
\hline
 \end{tabular}
 \label{tablepar}
\end{table*}

\begin{table*}
\caption{\small Observable quantities predicted from the sets of parameters in Table~\ref{tablepar}}
\label{tableobs}

\begin{tabular}{c|cccccccccccccc|c}
%\hline
Model  &  M/\msun  & $\chi^2_{{\rm R}i}$ &  $(Z/X)_{\rm s}$ & $\chi^2_{{\rm R}i}$ &  R/\rsol  & $\chi^2_{{\rm 
R}i}$ & $T_{\rm eff}$ & $\chi^2_{{\rm R}i}$ & L/\lsol  & $\chi^2_{{\rm R}i}$ &  $\Delta\nu$  & $\chi^2_{{\rm 
R}i}$ &  $\delta\nu$ & $\chi^2_{{\rm R}i}$ &$\chi^2_{\rm R}$\\
% mfix averaged_sep teff
\hline
\hline

A1t     & 1.105  &  & 0.037 &0.06 &  1.234 &  & 5764 &0.17 & 1.508 & 0.04  & 105.7 &0.01 & 4.95& 0.66  & 1.52 \\
B1t    & 0.934  &  & 0.039  &0.00&  0.874 &  & 5280 &0.03 & 0.534 & 0.44  & 161.1 & 0.& 9.22& 0.1  & \\
% mfix averaged_sep radii
\hline
A1r     & 1.105&  & 0.038&0.02  &  1.227& 0.20  & 5770& & 1.498& 0.13   & 106.5 &0.52 & 5.07& 0.53  & 2.80 \\
B1r     & 0.934&  & 0.041&0.00   &  0.872& 0.67  & 5289& & 0.533& 0.48   & 161.6 &0.11 &  9.33&0.14  &\\
% mfix averaged_sep no diff
\hline
A1nd     & 1.105&  &  0.040 & 0.00 & 1.227& 0.17  & 5780 & & 1.508& 0.05  &  106.6& 0.56  &  5.34& 0.27  & 
2.27\\
B1nd     & 0.934&  &  0.040 & 0.00 & 0.872& 0.66  & 5258 & & 0.521& 0.17  &  161.7& 0.13  &  9.60& 0.24 &\\
% mfix averaged_sep CEFF
 \hline
A1e     & 1.105  & & 0.038& 0.02  & 1.228& 0.24  & 5772&  & 1.500& 0.10  &  106.5& 0.46  &  5.16& 0.43  & 2.64\\
B1e     & 0.934  & & 0.041& 0.01  & 0.872& 0.66  & 5286&  & 0.533& 0.44  &  161.6& 0.44  &  9.42& 0.17 &\\
\hline
% mfix averaged_sep radii+atmo
%A1ra    & 1.105&   &  0.038&0.03   &  1.226& 0.06  & 5772 & &1.498& 0.14   & 106.2 &0.22  &  5.08& 0.51  & 
%2.10\\
%B1ra     & 0.934&  &  0.040&0.00 &  0.870& 0.44   & 5297 & &0.535 & 0.50    & 161.6 &0.44  &  9.3  &0.14 & \\
% \hline
% mfix rox + atmo
A1-3    & 1.105 &   &  0.035 &0.07   &  1.224& 0.00   & 5767 & &1.488& 0.16   & 106.5 &   &  5.77&  0.99  & 
2.08\\
B1-3    & 0.934&    &  0.038 &0.01   &  0.871& 0.29   & 5315 & &0.543 & 0.51  & 161.6 &0.04  &  9.85  &  & \\
\hline
% changing mass + averaged_sep
A2     & 1.104   & 0.00  & 0.039& 0.01  &  1.226& 0.13 & 5782 && 1.509& 0.04    & 106.6& 0.55  & 5.35& 0.26   & 
2.20 \\
B2      & 0.926  & 0.32 & 0.042& 0.01  &  0.870& 0.38 & 5259 && 0.520& 0.12    & 161.6& 0.09 & 9.72 & 0.30    & 
\\
\hline
% changing mass + averaged_sep + 1 MLT
A2c     & 1.107   & 0.01  & 0.036& 0.07  &  1.227& 0.22 & 5778 && 1.508& 0.04    & 106.6& 0.45  & 6.28& 0.03   & 
2.89 \\
B2c      & 0.919  & 0.99  & 0.040& 0.00  &  0.870& 0.37 & 5210 && 0.501& 0.00    & 161.0& 0.00 & 10.54 & 0.69    
& \\
\hline
% changing mass + averaged_sep + FST
A2f     & 1.105& 0.00    & 0.039 &0.01 &  1.226& 0.12  & 5784 && 1.510& 0.03    & 106.4& 0.32  & 5.35 & 0.27  & 
1.80 \\
B2f      & 0.927&0.28  & 0.041  &0.01 &  0.869& 0.28  & 5261 && 0.519& 0.13   & 161.6 & 0.07  & 9.72 & 0.29  & 
\\
\hline
% changing mass + averaged_sep + 1 punto SSEP B
%A2s     & 1.105& 0.00    & 0.039& 0.01  &  1.227& 0.20  & 5782 && 1.510& 0.03    & 106.5& 0.51  & 5.60 & 0.11  
%& 1.88 \\
%B2s      & 0.926& 0.38  & 0.042& 0.02 &  0.870& 0.39   & 5262 && 0.520& 0.15   & 161.6 & 0.07  & 10.26 & 0.01  
%& \\
%\hline
% rox mlt
A3    & 1.110& 0.06  &  0.038 & 0.01  & 1.224 & 0.00 & 5791 & &1.512& 0.01  &  106.6&  &  5.87  & 0.94&1.46\\
B3     & 0.927& 0.18  &  0.042& 0.02  & 0.869 & 0.16& 5259 & &0.518& 0.07  &  161.4 & 0.03 &  10.16 & &\\
 \hline
% rox 1 param mlt
A3c     & 1.113& 0.17  & 0.038& 0.02   & 1.223& 0.01  & 5819& & 1.540& 0.04 & 106.8 &  &  5.77 & 0.87& 1.67\\
B3c     & 0.922& 0.40  & 0.041& 0.00    & 0.870& 0.16  & 5206& & 0.497& 0.01 &  161.1& 0.00 & 10.22&   &\\
 \hline
% rox mlt sun
$\rm A3\alpha_{\odot}$ & 1.114 & 0.15   & 0.038 & 0.02  & 1.224& 0.00  & 5807&  & 1.529& 0.00  & 106.8 & & 5.94& 
0.75  & 1.60\\
$\rm B3\alpha_{\odot}$ & 0.921 & 0.47   & 0.041 & 0.01 & 0.869 & 0.15 & 5190&  & 0.491& 0.04 & 160.7 & 0.02  & 
10.35&  &\\
\hline
% rox fst atmos no bug
A3f     & 1.110 & 0.06 & 0.038& 0.01   & 1.224& 0.00  & 5790&  & 1.511&0.02  & 106.5&   & 5.85 & 0.92  & 1.40\\
B3f    & 0.927 & 0.16 & 0.042& 0.01  & 0.868 & 0.12 & 5260&  & 0.518&0.07  & 161.4& 0.02   & 10.14&   &\\
% rox no diff
\hline
A3nd     & 1.108& 0.02  &  0.040& 0.00 & 1.224& 0.00   & 5795& & 1.516& 0.01  &  106.5&  &  5.90 & 0.95 & 1.31\\
B3nd     & 0.929& 0.10  &  0.040& 0.00 & 0.869& 0.19  & 5240& & 0.511& 0.02  &  161.6& 0.04  &  10.16&  &\\
% rox ceff
\hline
A3e     & 1.109& 0.05  &  0.038& 0.01  & 1.224& 0.00  & 5792& & 1.514&0.01 & 106.6 &  & 5.89 & 0.95 & 1.45\\
B3e     & 0.927& 0.16 &  0.042 & 0.01  & 0.869& 0.17  & 5256& & 0.517&0.06  &  161.5&0.03  & 10.17 &  &\\
\hline
% rox fst
%A4     & 1.110& 0.07  & 0.038 & 0.01  & 1.224 & 0.00& 5790&  & 1.513& 0.01  & 106.3&   & 5.82& 0.94   & 1.38\\
%B4     & 0.927& 0.15  & 0.041 & 0.01 & 0.866 & 0.10& 5262& & 0.518& 0.07  & 161.4& 0.01   & 10.11&   &\\
% \hline

% rox fst atmos no bug + 2 sigma massB
%A4as     & 1.107 & 0.01 & 0.039& 0.01   & 1.224& 0.00  & 5797 &  & 1.520 &0.00  & 106.2&   & 5.91 & 0.92  & 
%1.17\\
%B4as     & 0.921 & 0.15 & 0.042& 0.01  & 0.866 & 0.05 & 5242&  & 0.509   &0.01  & 161.4 & 0.01   & 10.25&   &\\
% \hline

% rox overshoot
A4  & 1.112& 0.13  &  0.036& 0.05  & 1.224 &0.00&  5782& & 1.503& 0.05 &  106.7 &     & 5.99& 1.16 &1.96\\
B4  & 0.925& 0.27  &  0.039& 0.0   & 0.869 &0.15 & 5275& & 0.524&0.13  &  161.5& 0.02  & 10.56& &\\
% rox mix2
\hline
A5  & 1.109 & 0.04 &  0.028 & 0.02 & 1.224 & 0.00& 5791 & & 1.513 & 0.01 &  106.7&   &  5.86& 0.95  &1.51\\
B5  & 0.927 & 0.17 &  0.030 & 0.01 & 0.869 & 0.20 & 5251&  & 0.516& 0.05 &  161.6& 0.05  &  10.12 &    &\\

% no seismology
\hline
Ans     & 1.105 &  & 0.039 &  & 1.224 & & 5800 & &1.521  & &107.0 &  &  3.57 & & 0.06\\
Bns     & 0.934 &  & 0.041 &  & 0.863 & & 5238 & &0.502  & &164.3 &  &  8.27 & &\\
\hline

 \end{tabular}
\end{table*}

\subsection{Effect of classic and  seismic observable quantities}\label{sec:cal1}

A first calibration (A1r,B1r) was performed including in the
$\chi^2$ function  the luminosity, the radii and the
actual $(Z/X)_{\rm s}$ , as well as  the average large and small
frequency separations (computed as described in
Sect.~\ref{sec:cal_proc}) for each component. The model parameters
 are the initial chemical composition
($Y_0$,$Z_0$), the mixing length parameters ($\alpha_{\rm A}$,
$\alpha_{\rm B}$) and the age ($\tau$). In this calibration we
consider, following  \citet{Eggenberger04}, that the masses are
perfectly determined, and we assume the mass of each component to be 
fixed to its observational central value, as given 
\citet{Pourbaix02}. The parameters providing the
minimum $\chi^2$ are: $\tau=6.8 \pm 0.5$~Gyr, $\alpha_{\rm
A}$=2.00, $\alpha_{\rm B}$=2.24,  $Y_0$=0.276 and $Z_0$=0.0325. We
note that these results are in complete agreement with those
obtained by \citet{Eggenberger04}: $\tau=6.5\pm0.2$~Gyr, 
~$Z_0=0.0302$, ~$Y_0=0.275$, and also their value for the
mixing-length parameter of $\alpha$Cen~B  is $\sim$~10\% larger
than $\alpha_{\rm A}$.

This agreement is not surprising, as similar observational
constraints were considered. It strengthens the results obtained,
since  a different calibration procedure and different stellar
evolution codes (with different equation of state, treatment of
diffusion and treatment of sub-photospheric boundary conditions)
were used, and it provides us with a good reference model to study
the dependence of the calibrated stellar parameters on the choice
of observable quantities and of the physics.

The observational values of the masses given by \citet{Pourbaix02}, though precisely determined, should be
treated as observables and therefore introduced, with their error bars, in the definition of the $\chi^2$
and allowed to be changed during the calibration.
 $\rm M_A$ and $\rm M_B$ are considered both as parameters and observables in a second set of fittings
  (A2,B2; A2f,B2f). The readjustment  of parameters leads to a decrease of $M_{\rm B}$ which is  1.5$\sigma$
smaller than the value determined by \citet{Pourbaix02}, and to a decrease of age
 ($\tau=6.4\pm 0.6$~Gyr instead of $6.8 \pm 0.5$~Gyr). The location of $\alpha$Cen~B in the HR diagram
 has significantly improved compared with (A1r,B1r) (Fig.~\ref{fig:hr}) and $\langle\Delta\nu_{\rm A}\rangle$
  is also better reproduced (Fig.~\ref{fig:seismo1}), leading to an overall lower $\chi^2_{\rm R}$ compared with
  the equivalent fitting with fixed masses.

Even including the stellar masses among the parameters, we are not
able to improve significantly the fit of radii and large
separations. Actually the large separation is strongly dependent
on radius ($\Delta\nu\propto(M/R^3)^{1/2}$), and, given the high
precision of radius data, the procedure  privileges sets of
parameters providing the radii within 1$\sigma$ (for A), and
1.5$\sigma$ (for B) to detriment of a too high  large
separation: $\Delta\nu_{\rm A}$ is always around 106.6 $\mu Hz$
instead of 105.5 $\mu{\rm Hz}$. In order to relax  the
constraints, we have performed a fitting including \teff's among
the observables instead of the  radii (A1t,B1t). This fitting
provided a small $\chi^2_{\rm R}$ thanks to the good match of
large separation values. However, the radii (not included in the
$\chi^2$ function) are systematically larger (by more than 2$\sigma$)
than \citet{Kervella03} ones. The large separation is very much
affected by  the external layers properties, such as either the
description of the super-adiabatic region in the upper boundary of
the convective zone,  or the non-adiabatic processes (not taken
into account either in the stellar models or in the oscillation
code). An inspection of frequencies predicted for these models
(Fig.~\ref{fig:freq}) shows that  even if the fit of
$\Delta\nu_{\rm A}$ is almost perfect, the frequencies show a
shift of 25~$\mu{\rm Hz}$ with respect to the values determined by
\citet{Bouchy02}. On the other hand, sets of parameters
with a better fit of the radii reduce significantly the shift
of frequencies.

{\it The small separation for the A component (Fig.~\ref{fig:seismo1}) suggest that
seismic observables would privilege younger models, whereas one of the two values of  $\delta\nu_{\rm B}$ 
($n=23$)
and the classical observables tend to a high value of the age.}
In fact, in
our calibration (Ans,Bns) where only classical observable have
been taken into account for the fit, we obtain a very good
agreement for masses, radii, luminosity  and effective temperature
(not taken as observable) for both components, and the age is
$\tau=8.9 \pm 1.8$~Gyr.

%\subsection{Effect of seismic observables}

Though the  small separation averaged value of component A is well reproduced by our models, the
slope of $\delta\nu$ as a function of $\nu$ differs from the theoretical prediction.
 One could argue that this could be a consequence of taking as observable the average value  of $\delta\nu$.
 \citet[][ and references therein]{Roxburgh03}, show that  the
Tassoul asymptotic result gives a poor fit both to the small
separations  of stellar models
and to the observed values for the Sun, and that a better fit is
obtained by using the ratios of small and the large separations
($r_{02}(n)$). This ratio depends only on the inner phase shifts
which are determined solely by the interior structure of the star
and are uninfluenced by the unknown structure of the outer layers.

\begin{figure}
\resizebox{\hsize}{!}{\includegraphics[angle=-90]{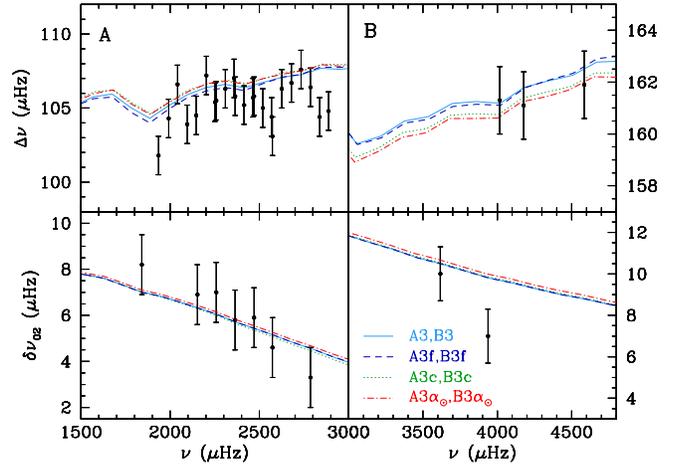}}
\caption
{\small As Fig.~\ref{fig:seismo1} but for models computed using different approach for convection:
MLT with two mixing length parameters (solid line); FST (dashed lines); MLT with $\alpha_{\rm A}=\alpha_{\rm B}$
  (dotted lines);
and MLT with $\alpha_{\rm A}=\alpha_{\rm B}=\alpha_{\odot}$ (dash-dotted lines)
Large and small separations are derived from stellar models, but the fitting is performed using  the ratios
$r_{02}$ for component A and $\Delta\nu$ for component B.}
\label{fig:seismo2}
\end{figure}

We have performed a similar fitting using six observational values of
the ratio $r_{02}$
 as seismic constraints for the A component. Unfortunately, the p-modes identified for
 B component do not allow us to define any value of $r_{02}$,   we decided, therefore,
 to take as seismic constraint the average large separation.
 The first thing to be noticed, when comparing the new set of parameters (A3,B3) with the one
 based on average large and small separation for both components (A2,B2), is a difference in the 
 resulting age of about 1~Gyr.
We also see that, by fitting $r_{02}$, we get a good fit of
$\delta\nu_{02}$ for both stars (Fig.~\ref{fig:seismo2}). The large separation is slightly
higher than the observational one, as we obtained in (A2,B2) and
(A1r,B1r).

In Fig.~\ref{fig:freq} we can see also the effect on the p-mode frequencies, the model
(A3,B3) providing better fit to the observational ones, than the model (A2,B2).

Notice that  the parameter set (A3,B3) was determined without
taking into account the small separation  for the
B component. Theoretical $\langle\delta\nu\rangle$ is $\sim 10.15\, \mu{\rm
Hz}$, that is close to the new observational value ($10.1 \mu{\rm Hz} $) by
 \citet{Kjeldsen04}. We wonder whether the high age obtained by 
taking $\langle\delta\nu_{\rm B}\rangle$ as given by the average of two points
\citep{Carrier03} is only a consequence of the low value
imposed to $\langle\delta\nu_{\rm B}\rangle$. In fact, a different calibration
 performed using the small
and large separations as observables, but taking only $\delta\nu_{\rm B}(n=21)$
instead of the average of two available values, provides an age of 6.0 Gyr, in
good agreement with $r_{02}$  (A3,B3) fittings, and quite younger
than (A2,B2) models.

Finally, we have also used $r_{02}$ but without varying the
masses. Again, the age is of the order of 6.0 Gyr, but now, the
mixing-length parameters needed for both stars are quite different
(by 17\%).

The fits reported in Table~\ref{tableobs} could indicate an  inconsistency
 between the seismic 
and classical observables. Actually for component B the resulting
radii are larger than the observed ones by more than $\sim 1\sigma$ or  $\sim
 2 \sigma$ and 
the masses are smaller by more than $\sim 1\sigma$ or  $\sim 2 \sigma$ with respect
 to the values observationally derived. 
These results could be interpreted as  a systematic error in the radii and/or in
 the mass determination.
A calibration performed with a free $M_{\rm B}$ parameter/observable, leads to $M_{\rm B}=0.911$~\msol\ and
$R_{\rm B}=0.863$~\rsol. The other parameters do not significantly change. This mass value would imply a systematic
error of $\sim 4\sigma$ that, given the available high precision data, does not seem reliable.
On the other hand, if $R_{\rm B}$ is let free, the fit of the other observables provides $R_{\rm B}=0.871$~\rsol\ instead
of 0.863\rsol, and a  $M_{\rm B}$ value   within $1\,\sigma$ from the observed one. 
It must be notice, however, that these fits  have included $\Delta\nu_{\rm B}$ in the $\chi^2$ functional.
To check if the large radius is only a consequence of $\Delta\nu_{\rm B}$, we calibrate the system adopting as seismic
constraints  $r_{02}(n)$ for component A, and $\delta\nu(n=21)$ for component B.   In this case we are able to fit
the radii and masses of both components within $1\sigma$, but $\langle\Delta\nu_{\rm B}\rangle$ is more than +3-$\sigma$ far away
from the observational value, and the predicted frequencies are $\sim$~40~$\mu{\rm Hz}$ larger than observed ones. 
A priori,  we  cannot rule out a larger  uncertainty in the observed radius. Actually  
in the observational radius determination there are implicit a definition of stellar radius (that is not 
necessarily the same than that used in stellar modelling) and an assumption about the limb darkening law and the atmosphere models.
 How much the stellar radius is affected by these assumptions?  Kervella (2005, private communication) claims that these effects
are much smaller that other errors intrinsic to the measurement method and already taken into account.
So, we should wait for more precise seismic data for $\alpha$CenB to understand this apparent inconsistency.

\subsection{Effect of the treatment of convection}
\label{sec:conv}
It is well known that one of the weak points of stellar evolution
models is the treatment of convection. The ``standard model'' of
convection adopted in stellar evolution is the mixing length
theory (MLT) where turbulence is described  by a relatively simple
model that contains essentially one adjustable parameter: the
mixing length $\Lambda=\alpha H_{\rm p}$ ($H_{\rm p}$ being the
local pressure scale height and $\alpha$ an unconstrained
parameter). The value of this parameter determines the radius of
the star and the behavior of the super-adiabatic region in the
outer boundary of the stellar convective zone. The calibration of
a stellar code using the solar radius and solar luminosity
provides the value of the mixing-length parameter
($\alpha_{\odot}$) which in turn is usually
 used to model other stars with the same
physics. A question arises: can stars of different masses, initial
chemical composition and evolutionary status be modeled with a
unique $\alpha$? If the answer is negative, how reliable is the
shape of the isochrones determined with a unique $\alpha$?

$\alpha$Cen offers  a unique opportunity of testing our
assumption about $\alpha$ and, therefore, our simplified  way of overcoming
the complex problem of convection in stars. The question of
the universality of the mixing-length parameter has been approached
many times in the past \citep{Noels91,Edmonds92,Lydon93,Neuforge93,Morel00,Fernandes95,Guenther00}. Some
attempts  of calibrating $\alpha$Cen~A\&B  in luminosity and radii
using MLT theory suggest different values of
$\alpha$ for each component and different from the
$\alpha_{\odot}$, while others favor similar values for both
components, or conclude that the uncertainties in masses and radii
as well as the chemical composition  should be reduced
significantly before being able to draw firm a conclusion on
whether the MLT parameter is unique or not \citep{Lydon93,Andersen91,Guenther00}. Nowadays, the high
precision with which we know masses and radii for $\alpha$Cen~A\&B
allows to analyze again this problem. The frequencies  are very
sensitive to $R$ and therefore to $\alpha$, and
 the difference between the observed  frequencies ($\nu_{\rm obs}$) and the theoretical ones
 ($\nu_{\rm theo}$) is highly affected by how the super-adiabatic zone is described
\citep[see e.g.][]{Schlattl97}.

  In order to analyze these questions we have made several fits of the $\alpha$Cen observable quantities
  by using MLT with  {\it i)}  $\alpha_{\rm A}\neq \alpha_{\rm B}$ as free parameters; {\it ii)}
  $\alpha_{\rm A} = \alpha_{\rm B}$  free parameter; and {\it iii)} $\alpha_{\rm A} = \alpha_{\rm B} = 
\alpha_{\odot}$
  fixed;  and {\it iv)} since \citet{Morel00} presented also controversial results when
  comparing calibrations for $\alpha$Cen using  MLT or FST, we shall analyze as well
  the effect of using the FST treatment of convection.

\subsubsection{MLT versus FST}
 Mimicking the spectral distribution of
eddies by one ``average'' eddy, such as MLT theory does, has
critical consequences on the computation of MLT fluxes. \citet{Canuto96} 
shows that in the limit of highly efficient convection MLT
underestimates the convective flux , and on the contrary, in the
low efficiency limit, MLT overestimates the convective flux. FST
models attempt to overcome the one-eddy 
approximation by using a
turbulence model to compute the full spectrum of a turbulent
convective flow. As consequence, the convective flux is $\sim$10
times larger than the MLT one in the limit of high efficiency, and
$\sim$0.3 in the limit of low efficiency. This behavior yields, in
the super-adiabatic region at the top of a convective zone, steeper
temperature gradients for FST than for solar-tuned MLT. The mixing
length used in FST treatment is in general defined as
 the distance from a given point to the
boundary of the convective zone. However, the FST fluxes have also
been used combined with other definitions of mixing length. For
instance, \citet{Bernkopf98}, use the FST fluxes with a mixing length
$\Lambda=\alpha^* H_{\rm p}$ with $\alpha^* < 1$. \citet{Morel00} 
use also this kind of description of FST convection with
CM91 for the fluxes. Here we will use this prescription of FST
treatment of convection, with the fluxes by \citet{Canuto96a}
instead of CM(91,92). The main difference is in the temperature gradient,
 steeper in CM than in CGM96, but both,  in any case, much steeper
than the MLT one. \citet{Morel00} found significant differences in
the fitted parameters (in particular a 1.5 Gyr difference in the
age) when investigating the effect of using a different theory of
convection in the modelling of $\alpha$ Centauri. 
This is rather surprising since the treatment of convection
affects only the very external layers of the star, and such an
important effect on the age of the star is not expected.

The convection parameters that result from our  fits are,
 as in the case of the calibration  with MLT,
 close to the parameter needed to calibrate the Sun (0.753). Furthermore,
 in comparing the models (A3,B3) with  (A3f,B3f) ones,  we
 do not observe any difference in the parameters of the models fitting the system $\alpha$Cen  using
 MLT or FST. In principle this is  a logic behavior since we considered as seismological observable
 the $r_{02}(n)$ \citep{Roxburgh03}  ratio, and this quantity
 is independent of surface properties of the model.
 However, also the sets of parameters (A2,B2) and (A2f,B2f) determined
  by  using large and small frequency separations as constraints,
 are in fact the same, independently of the convection description used.

{\it We would expect, like in the Sun, a decrease of the difference between
theoretical and observational frequencies in the high frequency
domain when FST is used.  However, for our binary system, the
improvement provided by FST treatment with respect to MLT is quite
small}.

 For $\alpha$Cen~A the improvement is of the order of 4~$\mu$Hz at 3000~$\mu$Hz, while
for the B component, this effect does not appear clearly (Fig.\ref{fig:freq}).
In fact, the
difference between MLT and FST frequencies  is due to a different 
radius and mass. If we adopt for $M_B$ and $R_B$ the values determined in the 
calibration B3 as the real ones, and compute  a new FST calibration, 
 we
find that the difference between the MLT and FST frequencies in the observational
domain (3000-4600$\mu$Hz) goes from 0 to 3~$\mu$Hz, while the difference in frequencies 
 between B3 and B3f introduced by the different mass and radius is of the order of 8~$\mu$Hz.

\subsubsection{ One or two different  values for the mixing length parameter?}

%\subsubsection{A different mixing length parameter for $\alpha$ Cen A and B?}
\label{sec:mlt}
Several times in the literature it has been addressed  the question if convection 
in the two components of  $\alpha$ Centauri should
 be described with distinct mixing-length parameters and, if, given the  observational uncertainties,
 the  inferred difference between mixing length parameters is significant
(see e.g \citet{Eggenberger04}, where an exhaustive review
  on previous calibrations is also presented).
It was, in fact, already suggested in \citet{Guenther00} that a reduction in the observational uncertainties
 and the inclusion of seismic constraints in the modelling would allow a more robust inference on the
  mixing-length parameters of $\alpha$Cen~A and B.
This is now the case, thanks to the detection of solar-like oscillations and to the precise determination
of radii. These are compatible with effective temperatures spectroscopically determined  and significantly
reduce the error box in the HR diagram.
As a general result of the calibrations presented in the previous sections we find that the mixing-length
 parameter of calibrated model A ($\alpha_{\rm A}$) is approximatively 5--10\% smaller than $\alpha_{\rm B}$.

Numerical simulations of convection has recently permitted to
carry out a calibration of mixing-length parameter
through the HR diagram.
A function $\alpha(T_{\rm eff},\log g)$ is determined in order to reproduce the step
of the specific entropy provided by the atmosphere hydrodynamic models.
 Both calibrations,
the one by \citet{Ludwig99} (based on 2-D simulations) and
the one by \citet{Trampedach04} (based on 3-D simulations) show slight
variations of $\alpha$ with the position in the HR diagram, and
suggest that the mixing parameter should be represented as a
function of
 $\log T_{\rm eff}$, $\log g$, and chemical composition.

{\it Actually,  comparing our results with these theoretical predictions, we see that  the difference
 between the
 value determined for $\alpha_{\rm B}$  and  $\alpha_{\rm A}$ is in good agreement with the
 predictions by numerical simulations of convection.} Moreover, our  $\alpha$ values  for the two
components  bracket  that 
 obtained by calibrating the Sun ($\alpha_\odot=1.91$) with same
physics (see for instance the model A3,B3). This is what  one
expects from their HR location and from the calibrations by
\citet{Ludwig99}  and \citet{Trampedach04}.
 We note, however, that
in the fits obtained by  using $\langle\Delta\nu\rangle$ and $ \langle\delta\nu\rangle$ 
we find $\alpha_{\rm B} > \alpha_{\rm A}$, but also  $\alpha_{\rm A}$ larger than the solar one. The difference with respect to the solar
value is even larger when the masses of the components are assumed to be fixed.
 The same effect is present in 
 \citet{Eggenberger04}: they find  $\alpha_{\rm B} > \alpha_{\rm A}$, but their 
values are far 
from their solar one.

In order to determine whether the addition of an extra free parameter leads to a significant improvement
 of the fit, we performed a calibration assuming a single mixing-length parameter for both components,
 and including the $r_{\rm 02}(n)$ ratios in the quality function
 (A3c,B3c).  The value  of $\alpha$ for this calibration is $\alpha=1.95$ (quite close to the solar one),
 but the masses now are slightly larger for the component A, and slightly smaller for the component
 B. The same happens if we decide to fix $\alpha_{\rm A}=\alpha_{\rm B}=\alpha_{\odot}$.
 The masses are anyway within 2$\sigma$ of the observed value, and the age is the same $\tau=5.7-5.8$
 Gyr.

 A different result is obtained if the small and large differences are considered in the
 quality function (A2c,B2c), and the mixing-length parameter is assumed to be the same for both
 stars. In this case, the value reached ($\alpha=1.78$) is not so close to the solar one,
 $M_{\rm B}$ must decrease to 0.919\msol, and the age of the system  decreases with respect to
 the value obtained allowing two different mixing-length parameters. This result recalls that
obtained by \citet{Thevenin02}, who using a unique $\alpha$ had to  decrease the
 $M_{\rm B}$  to 0.907\msol.

Since adding a free parameter would naturally lead to a better fit
(a lower $\chi^2$ as defined in Eq.(\ref{eq:chisq})), for a
quantitative comparison between the quality of the fit obtained
with a different
 number of model parameters, it is more meaningful to compare the so-called ``reduced'' $\chi^2_{\rm R}$ 
(Eq.~\ref{eq:redchisq}).
As can be seen in Table \ref{tableobs} the value of $\chi^2_R$ is
lower if two distinct mixing length parameters are used in the
modelling (both comparing A3,B3 with A3c,B3c and A2,B2 with A2c,B2c):
this suggests that, with the adopted observational constraints,
the addition of another free mixing-length parameter is justified.
In fact, comparing the fits obtained with one or two mixing-length parameters, the F-test gives a confidence of 
85-90\% that the inclusion of two different parameters for convection is significant. We should however recall 
that such a statistical test, and generally a $\chi^2$ statistics, assumes that the observational errors are 
distributed about the mean following a Gaussian distribution. This is not necessarily true as systematic shifts 
in the observed quantities and inaccuracies in the models cannot be excluded.
 As could also be expected, the uncertainties adopted with the observational
 constraints are crucial. For instance, we find that if the observational error in the  mass of the component B
  is doubled, the addition of a second free parameter for convection is no longer justified.

\begin{figure}
\resizebox{\hsize}{!}{\includegraphics[angle=-90]{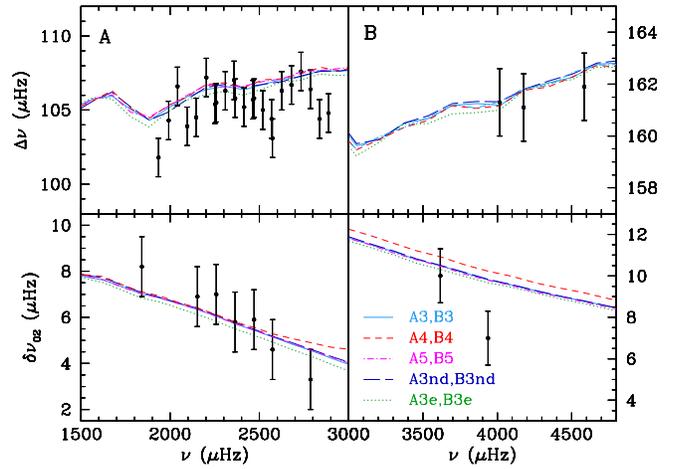}}
\caption{\small As Fig.~\ref{fig:seismo2} but with different curves corresponding to different physics
included in the stellar computation: convective overshooting (dashed lines);
solar mixture from  \cite{Asplund04} instead of \citet{Grevesse93}(dash-dotted lines); no
gravitational settling (solid lines);  CEFF equation of state instead of OPAL01 (dotted lines).
}\label{fig:seismo3}
\end{figure}

\subsection{Disentangling the physics}
\label{sec:phys}
 One of the principal motivations for pursuing the study of pulsations   in other stars is to test the
  assumptions concerning
 the physics underlying the stellar structure theory. To practically investigate  this idea, we compute
 four models of
  $\alpha$Cen  that  use the same observational constraints and model parameters
  as the reference calibrations (A3,B3) and (A1r,B1r), but incorporate
  changes in the physical assumptions. These are : CEFF EoS (A1e,B1e; A3e,B3e), \citet{Asplund04} 
chemical composition  (A5,B5). To test the effects
  of  gravitational settling we have calibrated the models (A1nd,B1nd) and (A3nd,B3nd)
  without microscopic diffusion.

\begin{figure}
\resizebox{\hsize}{!}{\includegraphics[angle=0]{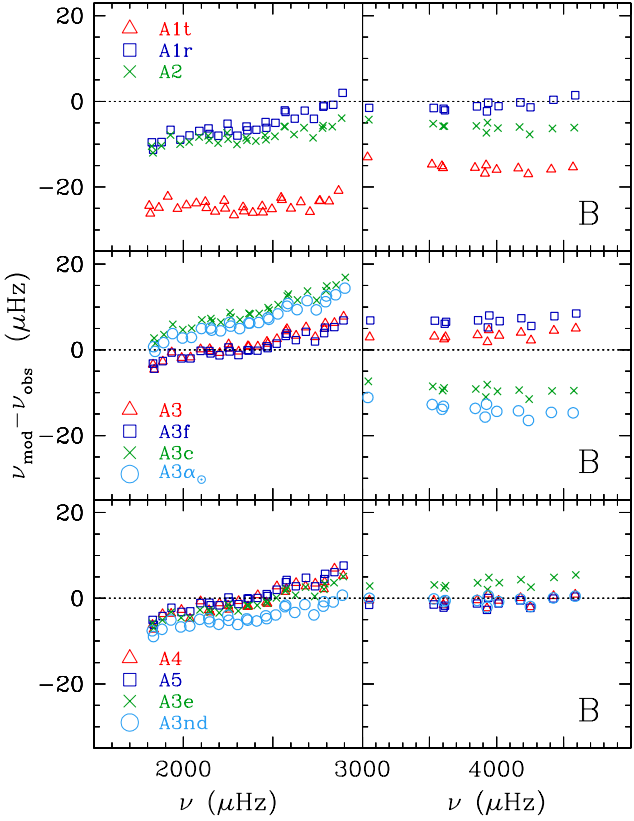}}
\caption{Difference between theoretical and observed frequencies for the sets of parameters whose
separations have been plotted in Fig.~\ref{fig:seismo1} (upper panel); in Fig.~\ref{fig:seismo2} (middle panel);
and Fig~\ref{fig:seismo3} (lower panel). Left side corresponding to $\alpha$Cen~A, and right side to
$\alpha$Cen~B. The labels correspond to the identification of the model used in Table~\ref{tablepar}.}
\label{fig:freq}
\end{figure}

\begin{figure}
\resizebox{\hsize}{!}{\includegraphics{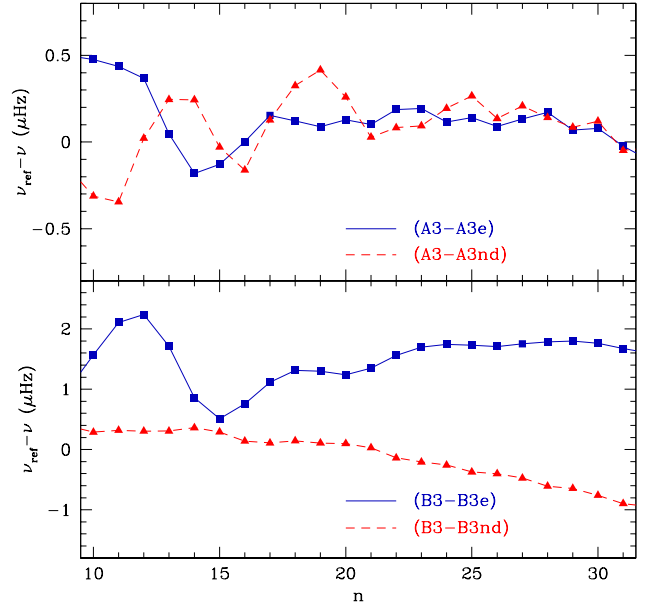}}
\caption{Difference of frequencies between a reference calibration (A3,B3), and  both, that with a different
EoS (A3e,B3e) (solid lines), and without microscopic diffussion (A3nd,B3nd) (dahsed lines). Upper panel 
corresponds to component A, and lower panel to component B.}
\label{dfreq}
\end{figure}

\begin{figure}
\resizebox{\hsize}{!}{\includegraphics[angle=-90]{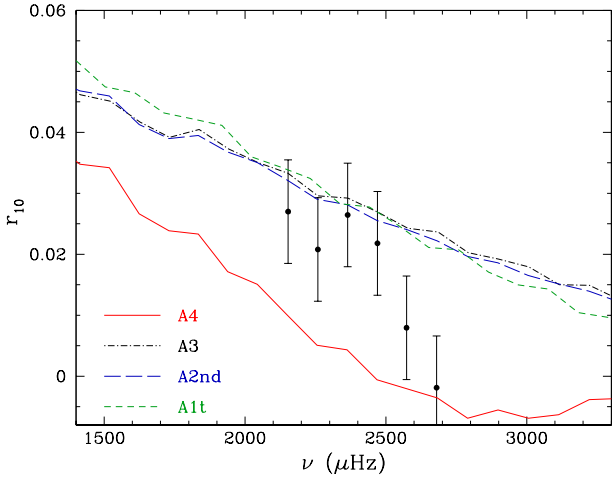}}
\caption{\small  $r_{10}(n)$ ratios for A component. Points represent the observational values with their 
error bars assuming an error in frequencies equal to $2\sigma$. Short-dashed line corresponds to the model calibrated
using large and small separation in the $\chi^2$ functional, as well as the effective temperatures instead of 
the
radii. The other three curves  correspond to modes calibrated using $r_{02}(n)$ in $\chi^2$:  model including
overshooting (solid line) and presenting  a convective core; the  same calibration without overshooting (dot-dashed line), and
calibration without overshooting and without microscopic diffusion (long-dashed line).}
\label{fig:rox}
\end{figure}

{\it   The central question is whether the residual errors are observationally significant. If the residuals
  are all small compared to the observational errors, it is not possible to distinguish parameter changes
  from changes in physical assumptions. In general, we find that stars obeying different physics
  succeed remarkably well in masquerading as stars that merely have different parameters} \citep{Brown94}.

\subsubsection{Equation of state}
Concerning the equation of state, the calibration assuming CEFF (A3e,B3e)
 leads to a result in agreement with the fit
(A3,B3) computed adopting OPAL01, both in terms of $\chi^2_{\rm R}$ (either the total one or that 
corresponding at each observable quantity), and of the fitted parameters. 
This is expected for stars with an internal structure similar to
the sun: as shown by \citet{Miglio04} the differences
between the sound speed and the first adiabatic exponent ($\Gamma_1$), due solely to
the use of a different EOS (CEFF and OPAL01), are smaller than 1\% except
in outer regions with radii larger than 0.95~R$_*$, that is
of the same order of those predicted in solar models  \citep[see e.g.][]{Basu97}.
The differences appear to be larger in the lower-mass model B than in model A,
and are mainly located in the hydrogen and helium ionization
regions. In that study, the differences in $\Gamma_1$ came from the application
of CEFF or OPAL01 to a given stellar structure ($\log \rho\, vs. \log T$) and chemical
composition. 
In the calibration procedure,  such small differences in the internal structure of a
model propagate in a variation of the observables of each model
that could be easily compensated by a re-adjustment of the free
parameters.

The models calibrated by using different  equations of state are very similar.
There is, nevertheless,  a difference in the frequencies. In Fig.~\ref{dfreq} (solid lines)
we plot the difference between (A3,B3) and (A3e,B3e) frequencies.
The upper panel corresponds to the A component, and the lower panel to the B one. 
This difference in  frequencies shows a behavior depending on $\nu$ with an oscillatory signature.
This  is expected for instance, when  the models have  either different  locations of the 
convective region boundaries, or a different behaviour of $\Gamma_1$.
The bottom of the convective zone in $\alpha$Cen~A is located at $x_{\rm zc}=R_{\rm cz}/R_{*}=0.707$ for
A3 calibration, and  $x_{\rm zc}=0.711$ for A3e model. 
Therefore, the oscillatory signature  is probably related to the changes in
the second helium ionization zone. The amplitude of this oscillation is linked to
the difference in  depth of the $\Gamma_1$ bump, and the period 
contains information about the location of the difference in stellar structure.
 For star A, the amplitude of the oscillation between the order $n=10$ and 15 is  
around  0.7~$\mu{\rm Hz}$, while for the B component that is almost 2~$\mu{\rm Hz}$.

\subsubsection{Microscopic diffusion}
The calibrations  without diffusion
(A1nd,B1nd) and (A3nd,B3nd)  provide  fits of similar quality   with respect to the 
corresponding calibrations including gravitational settling, respectively
(A1r,B1r) and (A3,B3).
 As reported in Table \ref{tablepar} the parameters resulting from the
calibrations differ, as expected, in the initial chemical composition.
The smaller mass fraction of He required in these calibrations implies also  
a  slight increase in the age of the system: $6.3\pm0.4$~Gyr for (A3nd,B3nd) versus
$5.8\pm0.2$~Gyr for (A3,B3) (and $7.1\pm0.6$~Gyr for (A1nd,B1nd) versus
$6.8\pm0.5$~Gyr for (A1r,B1r)).
The (A3nd,B3nd) calibration is consistent with the one obtained by \citet{Thoul03}. Since they use
only $\alpha$Cen~A frequencies, their age estimation is not affected by the low  $\delta\nu$(B) value.

The mixing-length parameters are also re-adjusted to fit the stellar radius, and 
also in this case $\alpha_{\rm B}$ is almost 8\% larger than $\alpha_{\rm A}$, and both 
bracket the mixing length parameter obtained for a Sun calibrated without microscopic diffusion 
($\alpha_{\odot}=1.84$).

In Fig.~\ref{dfreq} (dashed-line) we show how diffusion affects the frequencies. 
We plot the residuals 
defined as $\nu_{n0}({\rm A3})-\nu_{n0}({\rm A3nd})$ (upper panel) and  $\nu_{n0}({\rm B3})-\nu_{n0}({\rm 
B3nd})$
 (lower panel).
For component A the residuals as function of the order $n$ show an oscillatory behavior,
 reflecting the changes in envelope He abundance ($Y_{\rm s}$). Actually the calibrated model including
  diffusion has a superficial He abundance  $Y_{\rm s}({\rm A3})=0.245$, while model A3nd, without 
  gravitational settling has $Y_{\rm s}=0.270$.This difference implies
as well a different opacity and, therefore, a different location of the boundary of the convective zone
($X_{\rm cz}({\rm A3nd})=0.725$). To avoid changing the figure scale,
  the curve of residuals in the upper panel  has been shifted down by  $2.5 \mu{\rm Hz}$; we see that the  
amplitude of the oscillatory signal is around $0.5$~$\mu{\rm Hz}$  between $n=10$ and $n=20$.

For  component B the curve of residuals  shows a completely different behavior. Since $\alpha$Cen~B is less
massive than its companion,
the mass contained in its convective zone is much larger and, therefore, the effect of 
microscopic diffusion is much smaller. The value of $Y_{\rm s}$ for B3 is less than 4\% 
smaller than that for B3nd, while for the component A, the difference in $Y_{\rm s}$ between both calibrations 
(A3 and A3nd) is around 10\%.

\subsubsection{Solar mixture}
% \textbf{[MIX2]} 
\cite{Asplund04} have recently proposed a substantial revision of the abundance of
C N and O in the solar photosphere. Since the value of the surface
metallicity of $\alpha$Cen is observationally determined relative
to the solar heavy element abundance, the solar $(Z/X)_{\rm
s}=0.0177$ resulting from the new solar calibration 
 would lower $(Z/X)_{\rm
s}$ in $\alpha$Cen by $\sim$ 30\%. We thus calibrate our
models assuming as observational constraint $(Z/X)_{\rm s}=0.029$
(instead of 0.039)  and using in the computations OPAL opacity tables calculated with the 
solar mixture proposed by \cite{Asplund04}.
The results of this calibrations, models (A5,B5), present a
different initial chemical composition, but no other significant
deviation from the global parameters of models (A3,B3) is noticed.
 As it happens for the Sun, the $\alpha$CenA model calibrated with \cite{Grevesse93} has a
deeper convective region compared with the model A5, but the uncertainty in the observational
data is quite larger than in the Sun, and this difference can be masked by other choice
of parameters.

 \subsubsection {Overshooting}
An additional consequence of our simple way of describing
convection in stellar models, is the need to parameterize
convective overshooting.
 In general \citep[see e.g.][]{Schaller92} convective core overshooting is necessary
  to fit isochrones of open clusters for masses larger than a given critical mass. The problem
  is to determine this critical mass
and to describe the transition between no-overshooting
mass domain and overshooting mass domain. On the one hand, the mass
and effective temperature of $\alpha$Cen A are quite close to
solar values, and one could think that no overshooting should be
introduced. Nevertheless the chemical composition of $\alpha$Cen A
is different from the solar one and the evolution of a
convective core does not necessarily show the same
behaviour. The mass of $\alpha$Cen places it in the boundary
region between models with and without a convective core. We have
made several calibrations varying the thickness of the
overshooting layer $ov=\beta*(\min(r_{\rm cv}, H_{\rm p}(r_{\rm
cv})))$ (where $r_{\rm cv}$ is the radius of the convective core)
with $\beta$ from 0.0, 0.1, 0.15 and 0.2. In fact, for values of $\beta \leq
 0.15$ no convective core remains after the PMS, therefore the
parameters resulting from the fitting are not changed.

In Tables~\ref{tablepar} and ~\ref{tableobs} we report the set of
parameters  and observables corresponding to models (A4,B4)
calibrated with $\beta=0.2$. As should be expected, including
overshooting reduces the age of the system. 
Fig.~\ref{fig:seismo3} shows  the large and small separations.
We find a possible direct indicator
of a convective core in the behavior of the small separation of
component A at high frequencies ($\nu > 2500$~$\mu{\rm Hz}$).
 That is  even more evident in the signature left in $r_{\rm 02}$.
 However, given the uncertainty (1.3$\mu{\rm Hz}$) affecting the p-modes
involved in the highest frequency $r_{02}(n)$ (or $\delta\nu(n)$),
 it is not possible, based only on $r_{02}(n)$ or $\delta\nu_{02}$,
 to rule out   a convective core in $\alpha$Cen~A.
The ratio $r_{10}(n)=(\nu_{n0}-2 \nu_{n1} + \nu_{n+1,0})/(2 \Delta\nu_{0}(n+1))$, however, 
 is much more eloquent, as  shown in Fig.~\ref{fig:rox}, 
and {\it allows us to reject model A4:  current observational constraints seem  not to be 
 in favour of a convective core in $\alpha$Cen~A}.

\section{Summary and conclusions}

\label{conclu}
We propose a calibration of the binary  system $\alpha$Cen by means of Levenberg-Marquardt  minimization 
algorithm applying it simultaneously to classical (photometric, spectroscopic and astrometric) and 
seismic observables. The main features of this sort of algorithm make it an ideal tool for the aim of 
this work: to practically  analyze the effectiveness of oscillation frequencies in constraining stellar model 
parameters and stellar evolution physics, by using the p-modes identified by \cite{Bouchy02} ($\alpha$Cen~A) and 
by \cite{Carrier03}($\alpha$Cen~B).
Actually thanks to  its low computational cost, this algorithm allows to search for the best model in the full 
7-dimensional parameter space describing the binary system, varying both the set of stellar observables, and the 
physics included in stellar modelling.
 
As starting point we assumed the same observables than \cite{Eggenberger04}. In spite of the different EoS 
(MHD), diffusion
 treatment \citep[][ for five elements separately]{Richard96}, cross section of nuclear 
reactions (NACRE) and atmospheric boundary conditions, we derive a set of stellar parameters (A1r,B1r) in 
complete agreement with theirs. 

By comparison between  calibrations with different observables we make clear  that care has to be taken when using $\Delta\nu_0$  to constrain fundamental stellar parameters. 
Given the strong dependence  of $\Delta\nu_0$ on surface layers, and our poor understanding of the physics 
describing 
outer stellar regions, seeking a perfect agreement between observed and predicted $\Delta\nu_0$ may bias our 
results toward, for instance, inaccurate radii.
This is clear when comparing the models A1t and A1r;
in the former a perfect agreement with the observed
$\Delta\nu$ is reached but the radii (and the frequencies, see Fig.
\ref{fig:freq}) deviate significantly from their observational values.

An additional source of systematic error in the calibrations
concerns the value of the small frequency separation of component
B. Since the observational data are still rather poor one of the
two measured values, $\delta\nu_{\rm B}(n=23)$ leads to
a higher age than suggested by $\delta\nu_{\rm A}$.

The value of $\delta\nu_{\rm B}$ predicted in our calibration
(A3,B3) (where $\delta\nu_{\rm B}$ is not included in the
$\chi^2$) is in agreement with $\delta\nu_{\rm B}(n=21)$ and with
the very recent value published by \citet{Kjeldsen04}. This also suggests that sufficiently precise seismic
data of one star are sufficient to determine fundamental
parameters of the system. In fact additional calibrations, not
shown in Table~\ref{tablepar}, where no seismic constraints of component B are
included, lead to fitted parameters compatible with e.g. (A3,B3)
even though a large discrepancy between predicted and observed
frequencies of $\alpha$Cen B is observed.

We therefore propose to use $r_{\rm 02}$ as a reliable seismic
constraint to determine fundamental parameters of a star. The large
frequency separation could, nonetheless, provide a first estimate
of the mean density and, as shown e.g. in \cite{Gough90}, an
useful information on localized features in stellar interior once
more accurate determinations of solar-like oscillations will be
available.

Section~\ref{sec:conv} was devoted to the study of the
effects on the calibration of our uncertainties concerning stellar
convection. If the calibrations, e.g. (A3,B3), are performed
considering a free parameter describing convection in each
component ($\alpha_{\rm A}$, $\alpha_{\rm B}$) we find, as
\cite{Eggenberger04}, $\alpha_{\rm B}$ 5-11\%
higher than $\alpha_{\rm A}$. Differently from Eggenberger et al.
(2004) we find that $\alpha_{\rm A} < \alpha_{\odot} < \alpha_{\rm
B}$: this is of primary relevance when making statements
concerning the difference between $\alpha_{\rm A}$ and
$\alpha_{\rm B}$, otherwise we would also have to justify an even
bigger (not expected) difference between $\alpha_{\rm A}$ and
$\alpha_{\odot}$. We notice also that our inferred values of
$\alpha$ follow the same trend predicted by MLT parameter calibration based on 
2D  \citep{Ludwig99} and  3D \citep{Trampedach04} hydrodynamic atmosphere models.

In order to draw more firmed conclusions on the significance of
considering $\alpha_{\rm A}$ and  $\alpha_{\rm B}$ as free
parameters we repeated our calibrations (A3,B3) and (A2,B2)
assuming a single parameter for both components ((A3c,B3c) and
(A2c,B2c)). The fit necessarily improves when an
additional free parameter is introduced in the calibration, 
nevertheless we 
find, with the observational constraints we adopted, the
difference between $\alpha_{\rm A}$ and $\alpha_{\rm B}$ 
significant.
We note also that the value of $\alpha$ obtained in  (A3c,B3c) and (A2c,B2c) calibrations
is  quite close to the corresponding $\alpha_{\odot}$.

We also find 
 that, contrary to what was obtained by \citet{Morel00},   the use of a different theory of
convection (MLT or FST) in our models does not change the set of  parameters derived
from the fitting. The only effect of convection model is 
 a slight improvement in the fit of high frequencies when using FST (in particular for
A component).

The available seismic data are not in favour of a convective
core in $\alpha$ Cen A, moreover, the overshooting parameter  needed for a convective core to
persist after the PMS ($\beta=0.2$) appears to be too large for a
model of the mass and chemical composition of $\alpha$ Cen A \citep[see
e.g.][]{Demarque04}.

Finally, we find that stars obeying different physics  provide similar fits  to those
obtained with stars that merely have different parameters. As a consequence, we are not
able from  present data to discriminate neither between different EoS, neither between diffusion or not
diffusion models. Fig.~\ref{dfreq} shows, however, that expected precision from space missions would
allow to apply  inversion  techniques to frequency data.
Adding seismic observables has significantly improved the determination of the system fundamental parameters,
but more and more precise observations are needed to be able to extract information about the internal
structure of the stars.

\begin{acknowledgements}
A.M and J.M acknowledge financial support from the Prodex-ESA Contract 15448/01/NL/Sfe(IC). A.M. is also thankful to 
Teresa Teixeira for her useful suggestions.
\end{acknowledgements}
%*****************

%%%%%%%%%%%%%%%%%%
%  BIBLIOGRAPHY %%
%%%%%%%%%%%%%%%%%%
\bibliographystyle{aa}
\small
\bibliography{alphacen}

\begin{thebibliography}{64}
\expandafter\ifx\csname natexlab\endcsname\relax\def\natexlab#1{#1}\fi

\bibitem[{{Alexander} \& {Ferguson}(1994)}]{Alexander94}
{Alexander}, D.~R. \& {Ferguson}, J.~W. 1994, \apj, 437, 879

\bibitem[{{Andersen}(1991)}]{Andersen91}
{Andersen}, J. 1991, \aapr, 3, 91

\bibitem[{{Asplund} {et~al.}(2005){Asplund}, {Grevesse}, {Sauval}, {Allende
  Prieto}, \& {Blomme}}]{Asplund05}
{Asplund}, M., {Grevesse}, N., {Sauval}, A.~J., {Allende Prieto}, C., \&
  {Blomme}, R. 2005, \aap, 431, 693

\bibitem[{{Asplund} {et~al.}(2004){Asplund}, {Grevesse}, {Sauval}, {Allende
  Prieto}, \& {Kiselman}}]{Asplund04}
{Asplund}, M., {Grevesse}, N., {Sauval}, A.~J., {Allende Prieto}, C., \&
  {Kiselman}, D. 2004, \aap, 417, 751

\bibitem[{{B{\" o}hm-Vitense}(1958)}]{Bohm58}
{B{\" o}hm-Vitense}, E. 1958, Zeitschrift fur Astrophysics, 46, 108

\bibitem[{{Baglin} \& {The COROT Team}(1998)}]{Baglin98}
{Baglin}, A. \& {The COROT Team}. 1998, in IAU Symp. 185: New Eyes to See
  Inside the Sun and Stars, 301

\bibitem[{{Basu} \& {Christensen-Dalsgaard}(1997)}]{Basu97}
{Basu}, S. \& {Christensen-Dalsgaard}, J. 1997, \aap, 322, L5

\bibitem[{{Basu} {et~al.}(1997){Basu}, {Christensen-Dalsgaard}, {Chaplin},
  {Elsworth}, {Isaak}, {New}, {Schou}, {Thompson}, \& {Tomczyk}}]{Basuetal97}
{Basu}, S., {Christensen-Dalsgaard}, J., {Chaplin}, W.~J., {et~al.} 1997,
  \mnras, 292, 243

\bibitem[{{Bernkopf}(1998)}]{Bernkopf98}
{Bernkopf}, J. 1998, \aap, 332, 127

\bibitem[{{Bevington} \& {Robinson}(2003)}]{Bevington}
{Bevington}, P.~R. \& {Robinson}, D.~K. 2003, {Data Reduction and Error
  Analysis for the Physical Sciences} (McGraw-Hill)

\bibitem[{{Bouchy}(2002)}]{Bouchy02a}
{Bouchy}, F. 2002, in Astronomical Society of the Pacific Conference Series,
  474

\bibitem[{{Bouchy} \& {Carrier}(2002)}]{Bouchy02}
{Bouchy}, F. \& {Carrier}, F. 2002, \aap, 390, 205

\bibitem[{{Brown} {et~al.}(1994){Brown}, {Christensen-Dalsgaard},
  {Weibel-Mihalas}, \& {Gilliland}}]{Brown94}
{Brown}, T.~M., {Christensen-Dalsgaard}, J., {Weibel-Mihalas}, B., \&
  {Gilliland}, R.~L. 1994, \apj, 427, 1013

\bibitem[{{Burgers}(1969)}]{Burgers69}
{Burgers}, J.~M. 1969, {Flow Equations for Composite Gases} (Flow Equations for
  Composite Gases, New York: Academic Press, 1969)

\bibitem[{{Canuto}(1996)}]{Canuto96}
{Canuto}, V.~M. 1996, \apj, 467, 385

\bibitem[{{Canuto} {et~al.}(1996){Canuto}, {Goldman}, \&
  {Mazzitelli}}]{Canuto96a}
{Canuto}, V.~M., {Goldman}, I., \& {Mazzitelli}, I. 1996, \apj, 473, 550

\bibitem[{{Canuto} \& {Mazzitelli}(1991)}]{Canuto91}
{Canuto}, V.~M. \& {Mazzitelli}, I. 1991, \apj, 370, 295

\bibitem[{{Canuto} \& {Mazzitelli}(1992)}]{Canuto92}
{Canuto}, V.~M. \& {Mazzitelli}, I. 1992, \apj, 389, 724

\bibitem[{{Carrier} \& {Bourban}(2003)}]{Carrier03}
{Carrier}, F. \& {Bourban}, G. 2003, \aap, 406, L23

\bibitem[{{Caughlan} \& {Fowler}(1988)}]{Caughlan88}
{Caughlan}, G.~R. \& {Fowler}, W.~A. 1988, Atomic Data and Nuclear Data Tables,
  40, 283

\bibitem[{{Chmielewski} {et~al.}(1992){Chmielewski}, {Friel}, {Cayrel de
  Strobel}, \& {Bentolila}}]{Chmielewski92}
{Chmielewski}, Y., {Friel}, E., {Cayrel de Strobel}, G., \& {Bentolila}, C.
  1992, \aap, 263, 219

\bibitem[{{Christensen-Dalsgaard} {et~al.}(1995){Christensen-Dalsgaard},
  {Bedding}, {Houdek}, {Kjeldsen}, {Rosenthal}, {Trampedach}, {Monteiro}, \&
  {Nordlund}}]{JCD95}
{Christensen-Dalsgaard}, J., {Bedding}, T.~R., {Houdek}, G., {et~al.} 1995, in
  Astronomical Society of the Pacific Conference Series, 447

\bibitem[{{Christensen-Dalsgaard} \& {D\"appen}(1992)}]{JCD92}
{Christensen-Dalsgaard}, J. \& {D\"appen}, W. 1992, \aapr, 4, 267

\bibitem[{{Christensen-Dalsgaard} {et~al.}(1996){Christensen-Dalsgaard},
  {Dappen}, {Ajukov}, {Anderson}, {Antia}, {Basu}, {Baturin}, {Berthomieu},
  {Chaboyer}, {Chitre}, {Cox}, {Demarque}, {Donatowicz}, {Dziembowski},
  {Gabriel}, {Gough}, {Guenther}, {Guzik}, {Harvey}, {Hill}, {Houdek},
  {Iglesias}, {Kosovichev}, {Leibacher}, {Morel}, {Proffitt}, {Provost},
  {Reiter}, {Rhodes}, {Rogers}, {Roxburgh}, {Thompson}, \& {Ulrich}}]{JCD96}
{Christensen-Dalsgaard}, J., {Dappen}, W., {Ajukov}, S.~V., {et~al.} 1996,
  Science, 272, 1286

\bibitem[{{Cox} \& {Giuli}(1968)}]{Cox}
{Cox}, J.~P. \& {Giuli}, R.~T. 1968, {Principles of Stellar Structure} (Gordon
  and Breach)

\bibitem[{{Demarque} {et~al.}(2004){Demarque}, {Woo}, {Kim}, \&
  {Yi}}]{Demarque04}
{Demarque}, P., {Woo}, J., {Kim}, Y., \& {Yi}, S.~K. 2004, \apjs, 155, 667

\bibitem[{{Di Mauro} {et~al.}(2003){Di Mauro}, {Christensen-Dalsgaard},
  {Kjeldsen}, {Bedding}, \& {Patern{\` o}}}]{DiMauro03}
{Di Mauro}, M.~P., {Christensen-Dalsgaard}, J., {Kjeldsen}, H., {Bedding},
  T.~R., \& {Patern{\` o}}, L. 2003, \aap, 404, 341

\bibitem[{{Edmonds} {et~al.}(1992){Edmonds}, {Cram}, {Demarque}, {Guenther}, \&
  {Pinsonneault}}]{Edmonds92}
{Edmonds}, P., {Cram}, L., {Demarque}, P., {Guenther}, D.~B., \&
  {Pinsonneault}, M.~H. 1992, \apj, 394, 313

\bibitem[{{Eggenberger} {et~al.}(2004){Eggenberger}, {Charbonnel}, {Talon},
  {Meynet}, {Maeder}, {Carrier}, \& {Bourban}}]{Eggenberger04}
{Eggenberger}, P., {Charbonnel}, C., {Talon}, S., {et~al.} 2004, \aap, 417, 235

\bibitem[{{Fernandes} \& {Neuforge}(1995)}]{Fernandes95}
{Fernandes}, J. \& {Neuforge}, C. 1995, \aap, 295, 678

\bibitem[{{Flannery} \& {Ayres}(1978)}]{Flannery78}
{Flannery}, B.~P. \& {Ayres}, T.~R. 1978, \apj, 221, 175

\bibitem[{{Gough}(1990)}]{Gough90}
{Gough}, D.~O. 1990, Lecture Notes in Physics, Berlin Springer Verlag, 367, 283

\bibitem[{{Grevesse} \& {Noels}(1993)}]{Grevesse93}
{Grevesse}, N. \& {Noels}, A. 1993, in La formation des \'el\'ements chimiques,
  AVCP, ed. R.~D. Hauck~B., Paltani~S.

\bibitem[{{Guenther} \& {Brown}(2004)}]{Guenther04}
{Guenther}, D.~B. \& {Brown}, K.~I.~T. 2004, \apj, 600, 419

\bibitem[{{Guenther} \& {Demarque}(2000)}]{Guenther00}
{Guenther}, D.~B. \& {Demarque}, P. 2000, \apj, 531, 503

\bibitem[{{Iglesias} \& {Rogers}(1996)}]{Iglesias96}
{Iglesias}, C.~A. \& {Rogers}, F.~J. 1996, \apj, 464, 943

\bibitem[{{Kervella} {et~al.}(2003){Kervella}, {Th{\' e}venin}, {S{\'
  e}gransan}, {Berthomieu}, {Lopez}, {Morel}, \& {Provost}}]{Kervella03}
{Kervella}, P., {Th{\' e}venin}, F., {S{\' e}gransan}, D., {et~al.} 2003, \aap,
  404, 1087

\bibitem[{{Kjeldsen} \& {Bedding}(1995)}]{Kjeldsen95}
{Kjeldsen}, H. \& {Bedding}, T.~R. 1995, \aap, 293, 87

\bibitem[{{Kjeldsen} \& {Bedding}(2004)}]{Kjeldsen04}
{Kjeldsen}, H. \& {Bedding}, T.~R. 2004, in ESA SP-559: SOHO 14 Helio- and
  Asteroseismology: Towards a Golden Future, 101

\bibitem[{{Kurucz}(1998)}]{Kurucz98}
{Kurucz}, R.~L. 1998, \texttt{http://kurucz.harvard.edu/grids.html}

\bibitem[{{Ludwig} {et~al.}(1999){Ludwig}, {Freytag}, \& {Steffen}}]{Ludwig99}
{Ludwig}, H., {Freytag}, B., \& {Steffen}, M. 1999, \aap, 346, 111

\bibitem[{{Lydon} {et~al.}(1993){Lydon}, {Fox}, \& {Sofia}}]{Lydon93}
{Lydon}, T.~J., {Fox}, P.~A., \& {Sofia}, S. 1993, \apj, 413, 390

\bibitem[{{Matthews}(1998)}]{Matthews98}
{Matthews}, J.~M. 1998, in Structure and Dynamics of the Interior of the Sun
  and Sun-like Stars SOHO 6/GONG 98 Workshop Abstract, June 1-4, 1998, Boston,
  Massachusetts, p. 395, 395

\bibitem[{{Metcalfe}(2005)}]{Metcalfe05}
{Metcalfe}, T.~S. 2005, ArXiv Astrophysics e-prints \texttt{[astro-ph/0501421]}

\bibitem[{{Miglio}(2004)}]{Miglio04}
{Miglio}, A. 2004, in AIP Conference Proceedings, Vol. 731, Equation-of-State
  and Phase-Transition in Models of Ordinary Astrophysical Matter, 187--192

\bibitem[{{Morel} {et~al.}(2000){Morel}, {Provost}, {Lebreton}, {Th{\'
  e}venin}, \& {Berthomieu}}]{Morel00}
{Morel}, P., {Provost}, J., {Lebreton}, Y., {Th{\' e}venin}, F., \&
  {Berthomieu}, G. 2000, \aap, 363, 675

\bibitem[{{Neuforge}(1993)}]{Neuforge93}
{Neuforge}, C. 1993, \aap, 268, 650

\bibitem[{{Neuforge-Verheecke} \& {Magain}(1997)}]{Neuforge-Verheecke97}
{Neuforge-Verheecke}, C. \& {Magain}, P. 1997, \aap, 328, 261

\bibitem[{{Noels} {et~al.}(1991){Noels}, {Grevesse}, {Magain}, {Neuforge},
  {Baglin}, \& {Lebreton}}]{Noels91}
{Noels}, A., {Grevesse}, N., {Magain}, P., {et~al.} 1991, \aap, 247, 91

\bibitem[{{Pijpers}(2003)}]{Pijpers03}
{Pijpers}, F.~P. 2003, \aap, 400, 241

\bibitem[{{Pourbaix} {et~al.}(2002){Pourbaix}, {Nidever}, {McCarthy}, {Butler},
  {Tinney}, {Marcy}, {Jones}, {Penny}, {Carter}, {Bouchy}, {Pepe}, {Hearnshaw},
  {Skuljan}, {Ramm}, \& {Kent}}]{Pourbaix02}
{Pourbaix}, D., {Nidever}, D., {McCarthy}, C., {et~al.} 2002, \aap, 386, 280

\bibitem[{{Richard} {et~al.}(1996){Richard}, {Vauclair}, {Charbonnel}, \&
  {Dziembowski}}]{Richard96}
{Richard}, O., {Vauclair}, S., {Charbonnel}, C., \& {Dziembowski}, W.~A. 1996,
  \aap, 312, 1000

\bibitem[{{Rogers} \& {Nayfonov}(2002)}]{Rogers02}
{Rogers}, F.~J. \& {Nayfonov}, A. 2002, \apj, 576, 1064

\bibitem[{{Roxburgh} \& {Vorontsov}(2003)}]{Roxburgh03}
{Roxburgh}, I.~W. \& {Vorontsov}, S.~V. 2003, \aap, 411, 215

\bibitem[{{S{\" o}derhjelm}(1999)}]{Soderhjelm99}
{S{\" o}derhjelm}, S. 1999, \aap, 341, 121

\bibitem[{{Salpeter}(1954)}]{Salpeter54}
{Salpeter}, E.~E. 1954, Australian Journal of Physics, 7, 373

\bibitem[{{Schaller} {et~al.}(1992){Schaller}, {Schaerer}, {Meynet}, \&
  {Maeder}}]{Schaller92}
{Schaller}, G., {Schaerer}, D., {Meynet}, G., \& {Maeder}, A. 1992, \aaps, 96,
  269

\bibitem[{{Schlattl} {et~al.}(1997){Schlattl}, {Weiss}, \&
  {Ludwig}}]{Schlattl97}
{Schlattl}, H., {Weiss}, A., \& {Ludwig}, H.-G. 1997, \aap, 322, 646

\bibitem[{{Tassoul}(1980)}]{Tassoul80}
{Tassoul}, M. 1980, \apjs, 43, 469

\bibitem[{{Th{\' e}venin} {et~al.}(2002){Th{\' e}venin}, {Provost}, {Morel},
  {Berthomieu}, {Bouchy}, \& {Carrier}}]{Thevenin02}
{Th{\' e}venin}, F., {Provost}, J., {Morel}, P., {et~al.} 2002, \aap, 392, L9

\bibitem[{{Thoul} {et~al.}(2003){Thoul}, {Scuflaire}, {Noels}, {Vatovez},
  {Briquet}, {Dupret}, \& {Montalban}}]{Thoul03}
{Thoul}, A., {Scuflaire}, R., {Noels}, A., {et~al.} 2003, \aap, 402, 293

\bibitem[{{Thoul} {et~al.}(1994){Thoul}, {Bahcall}, \& {Loeb}}]{Thoul94}
{Thoul}, A.~A., {Bahcall}, J.~N., \& {Loeb}, A. 1994, \apj, 421, 828

\bibitem[{{Trampedach}(2004)}]{Trampedach04}
{Trampedach}, R. 2004, Ph.D.~Thesis

\bibitem[{{Turcotte} {et~al.}(1998){Turcotte}, {Richer}, {Michaud}, {Iglesias},
  \& {Rogers}}]{Turcotte98}
{Turcotte}, S., {Richer}, J., {Michaud}, G., {Iglesias}, C.~A., \& {Rogers},
  F.~J. 1998, \apj, 504, 539

\end{thebibliography}
%\begin{thebibliography}{}
\end{document}